\documentclass[3p]{elsarticle}
\biboptions{sort&compress}

\usepackage{amsmath,amsfonts,amssymb,amsthm}
\usepackage{mathtools,bm,esint,physics,stmaryrd}
\usepackage{empheq,esdiff,esvect,yhmath,leftidx}
\MakeRobust{\vv}

\makeatletter
\@ifpackageloaded{esvect}{%
}{%
  \RequirePackage{amsmath}%
}
\makeatother

\usepackage{graphicx,float,placeins,booktabs}
\usepackage{longtable,multirow,array,tabu}
\usepackage[export]{adjustbox}
\usepackage{epstopdf}
\usepackage[T1]{fontenc}
\usepackage[utf8]{inputenc}
\usepackage[english]{babel}
\usepackage[font=footnotesize]{caption}
\usepackage[shortlabels]{enumitem}
\usepackage{xcolor}
\usepackage{siunitx}
\usepackage{stackengine,gensymb,framed,setspace}
\usepackage[normalem]{ulem}

\usepackage{algorithm,algpseudocode}

\usepackage{tikz}
\usetikzlibrary{positioning,arrows.meta,fit,backgrounds}

\usepackage{natbib}
\usepackage[hidelinks]{hyperref}   %
\usepackage{cleveref}

\usepackage{etoolbox}
\preto\subequations{\ifhmode\unskip\fi}

\usepackage{geometry}
\def\nten#1{\mathbf{#1}}

\def\lch{\text{l}_{\text{ch}}}
\def\Dbe{\myvect{\Delta \beta^\eps}}

\def\sts{\sigma_{\mathtt{ts}}}
\def\scs{\sigma_{\mathtt{cs}}}

\def\de{\delta^\varepsilon}
\def\ce{c_{\mathtt{e}}}

\def\sigmTwo{\sigma_{\mathtt{m,2}}}
\def\smTwo{s_{\mathtt{m,2}}}
\def\sigmax{\sigma_{\mathtt{max}}}
\def\sigmin{\sigma_{\mathtt{min}}}

\def\toct{\tau_{\mathtt{oct}}}

\def\bfe{{\bf e}}        
        
    \def\bfu{{\bf u}}    
    \def\bfx{{\bf x}}

    \def\bfE{{\bf E}}    \def\bfF{{\bf F}}
    \def\bfI{{\bf I}}    
        \def\bfN{{\bf N}}
        
\def\bfS{{\bf S}}        
    \def\bfX{{\bf X}}

\def\eps{\varepsilon}

\DeclareMathOperator*{\argmin}{arg\,min}
\DeclareMathAlphabet{\mathsfit}{T1}{\sfdefault}{\mddefault}{\sldefault}
\SetMathAlphabet{\mathsfit}{bold}{T1}{\sfdefault}{\bfdefault}{\sldefault}

\newcommand{\FNN}{F_{\mathtt{NN}}}
\newcommand{\Fhb}{F_{\mathtt{HB}}}
\newcommand{\Fmc}{F_{\mathtt{MC}}}
\newcommand{\Fmgc}{F_{\mathtt{MgC}}}
\newcommand{\Fdp}{F_{\mathtt{DP}}}

\newcommand{\stsNN}{\sigma_{\mathtt{ts}}^{\mathtt{NN}}}

\newcommand{\softplus}{\mathrm{softplus}}

\newcommand{\Itriax}{\eta}
\newcommand{\Lode}{\theta}

\newcommand{\devop}{\mathrm{dev}}

\newcommand{\myvect}[1]{\vv{#1}}
\newcommand{\mymatrix}[1]{\mathsfit{#1}}

\newcommand{\runin}[1]{\elsparagraph*{#1:}}

\newcommand{\etabar}{\bar{\eta}}                         %
\newcommand{\phihat}{\hat{\phi}}                         %
\newcommand{\pht}{p_{\mathtt{t}}}                        %
\newcommand{\phc}{p_{\mathtt{c}}}
\newcommand{\pbar}{\bar{p}}                              %
\newcommand{\etacrt}{\eta^{\mathtt{t}}_{\mathtt{cr}}}    %
\newcommand{\etacrc}{\eta^{\mathtt{c}}_{\mathtt{cr}}}    %
\newcommand{\shts}{\sigma_{\mathtt{hts}}}                %
\newcommand{\shcs}{\sigma_{\mathtt{hcs}}}                %
\newcommand{\ReLU}{\mathrm{ReLU}}
\newcommand{\clip}{\mathrm{clip}}

\newcommand{\mf}{\text{m}_{\text{f}}}

\theoremstyle{plain}
\newtheorem{theorem}{Theorem}

\newtheorem{remark}[theorem]{Remark}

\long\def\symbolfootnote[#1]#2{\begingroup%
\def\thefootnote{\fnsymbol{footnote}}\footnote[#1]{#2}\endgroup}

\begin{document}
\begin{frontmatter}

\title{Machine learning of strength surfaces enables high regularization length phase-field fracture\vspace{0.1cm}}

\author[purdue]{S. Chockalingam\corref{cor1}}
\ead{csenthil@purdue.edu}

\author[gatech,GTmech]{Aditya Kumar}
\ead{aditya.kumar@ce.gatech.edu}

\author[columbia]{Adrian Buganza Tepole}
\ead{ab6035@columbia.edu}

\address[purdue]{Department of Mechanical Engineering, Purdue University, West Lafayette, IN 47907, USA}
\address[gatech]{School of Civil and Environmental Engineering, Georgia Institute of Technology, Atlanta, GA 30332, USA}
\address[GTmech]{George W. Woodruff School of Mechanical Engineering, Georgia Institute of Technology, Atlanta, GA 30332, USA \vspace{0.05cm}}
\address[columbia]{Department of Mechanical Engineering, Columbia University, New York, NY 10027, USA}

\cortext[cor1]{Corresponding author}

\begin{abstract}
Classical phase-field models of brittle fracture describe crack growth governed by toughness, but they do not include the strength surface of the material, which controls the nucleation of fracture in the absence of pre-existing cracks. The phase-field theory of Kumar et al.~(2020) incorporates the material strength surface through a nucleation driving force. Chockalingam~(2025) subsequently derived a general expression for this driving force for arbitrary strength surfaces characterized by $n$ material parameters. The resulting phase-field strength surface (PFSS) converges to the prescribed material strength surface as the regularization length $\eps\to0$ and coincides with it exactly at $n$ selected strength states for any $\eps$. In practice, however, this approach requires prescribing an analytic form for the strength surface and calibrating its parameters against experimental data, a process that can be challenging when the underlying functional form is unknown and the available strength measurements are sparse, noisy, or incomplete, particularly for new materials. Moreover, away from the $n$ calibrated states, the PFSS deviates from the material strength surface at finite $\eps$, and for the widely used analytic strength surfaces, with only $n=2$, accurate strength reproduction requires $\eps$ to be commensurate with the Irwin characteristic length $\lch$, which imposes prohibitive mesh requirements for large structures. To address both limitations, we introduce a neural-network (NN) strength function for isotropic materials that is learned directly from strength data, is physics-constrained by construction, and provides far more calibration parameters than the analytic surfaces. The calibration is further generalized to admit any number of calibration strength states and to enforce openness of the PFSS in chosen directions. A single network architecture captures synthetic strength data generated from the Mohr--Coulomb, Drucker--Prager, Mogi--Coulomb and 3D Hoek--Brown surfaces, as well as experimental strength data for graphite, natural rubber and styrene--butadiene rubber. The resulting PFSS matches the material strength surface accurately up to regularization lengths as large as $\eps=10\,\lch$, whereas the PFSS of the exact analytic surfaces used to generate the synthetic data performs comparatively poorly even at $\eps=0.5\,\lch$. Finite-element simulations of canonical fracture problems validate the phase-field theory with the NN strength function across fracture regimes, capturing (i) nucleation under uniform stress, (ii) crack growth from large pre-existing flaws, and (iii) fracture governed jointly by strength and toughness. By decoupling strength fidelity from the regularization length, the present approach substantially reduces the mesh resolution required for phase-field fracture simulations of large structures, particularly in three dimensions.
\end{abstract}


\end{frontmatter}

\section{Introduction}\label{sec:intro}

Describing fracture in elastic brittle materials requires three material ingredients: elasticity, toughness, and strength~\cite{KBFLP20}. A unified mathematical theory incorporating all three remains to be fully developed, although substantial progress has recently been made in this direction~\cite{lopez2026nucleation}. Meanwhile, many modeling approaches have been put forward in its place over the past six decades to integrate the three ingredients robustly. The phase-field approach to fracture~\cite{Francfort98,Bourdin00}, which regularizes sharp cracks over a regularization length $\eps$, was originally developed to predict when and where cracks grow and has more recently emerged as a framework in which the three ingredients can be integrated systematically in a single formulation that is both broadly applicable and computationally viable. Of the models proposed within this framework, that of Kumar et al.~\cite{KFLP18,KBFLP20} has been demonstrated to be successful at describing fracture nucleation and propagation across a broad range of loading conditions, geometries and materials, including soft nonlinear materials~\cite{KLP21,KRLP22,LK24,KKLP24,KDK2025Comparison,Chockalingam2026EFM}.\\

The model of Kumar et al.~(2018)~\cite{KFLP18} was motivated by the observation that classical energy-based phase-field models can account for, at best, the uniaxial tensile strength of a material. As set out in~\cite{KFLP18, KBFLP20}, however, strength is not a single tensile value but a surface in stress space: the set of critical stresses at which a large homogeneous specimen fails under arbitrary uniform multiaxial loading. To incorporate this strength surface into the phase-field theory, Kumar et al.\ added a driving force $\ce$ to the partial differential equation that governs the evolution of the phase-field variable. When constructed appropriately, $\ce$ captures the strength surface while preserving the ability of the classical phase-field regularization to describe crack propagation according to Griffith's fracture postulate~\cite{KDK2025Comparison}. Subsequently, Chockalingam~(2025)~\cite{chockalingam2025construction}, following the construction laid out in \cite{KBFLP20}, derived a general explicit expression for $\ce$ that incorporates arbitrary strength surfaces $F(\bm{\sigma};\myvect{\beta})=0$ into the phase-field theory, and showed that the resulting phase-field strength surface (PFSS) exactly matches the material strength surface in the limit of vanishing regularization length $\eps$. He further showed that the PFSS can be made to match the material strength surface exactly at any $n$ independent calibration strength states, where $n$ is the number of material parameters $\myvect{\beta}$ describing the strength surface, irrespective of the value of $\eps$. This exact calibration is critical for capturing the strength surface accurately with the finite regularization lengths that are necessary in practice. Building on this, Chockalingam et al.~(2026)~\cite{Chockalingam2026EFM} demonstrated the incorporation of diverse and widely used analytic strength surfaces, namely the Mohr--Coulomb (M--C), 3D Hoek--Brown (3D H--B) and Mogi--Coulomb (Mg--C) surfaces, within a finite-element framework. Through canonical fracture tests, they demonstrated the ability of the formulation to capture fracture nucleation under uniform stress, crack propagation from large pre-existing cracks, and the transition between strength- and toughness-controlled fracture. These developments made it possible, in principle, to incorporate any strength surface into the phase-field theory.\\

Two practical limitations nevertheless remain. The first concerns the description of the strength surface itself. For a given material with measured strength data, an appropriate analytic form must be selected and its parameters calibrated. This is a manual, case-by-case exercise that limits the strength surfaces that can be modeled to the shapes that the available analytic forms can take, and makes calibration to experimental strength data cumbersome, particularly for new materials whose strength surface is not known a priori. The appropriate form is often unclear and may differ between materials of a similar class. Experimental strength data are also typically noisy and available only over a limited region of stress space. Moreover, commonly used analytic strength surfaces may fail to capture the failure behavior of certain materials, such as elastomers that exhibit non-convex strength surfaces~\cite{hamdi2006rubber}. For example, specialized functional forms have been developed to describe the strength surfaces of natural rubber (NR) and styrene--butadiene rubber (SBR)~\cite{hamdi2006rubber,rosendahl2019equivalent}. A promising alternative is to learn the strength function directly from experimental data, without prescribing its analytic form a priori, as has been explored extensively for yield surfaces in data-driven plasticity~\cite{Hartmaier2020,Vlassis2021,Nascimento2023}.\\

The second practical limitation, which is critical for modeling large-scale structures, is the size of the regularization length $\eps$ that can be used. The finite-element mesh in fracturing regions must resolve $\eps$, so that the number of unknowns, and hence the computational cost, grows rapidly as $\eps$ is decreased, particularly in three-dimensional simulations. Two considerations limit how large $\eps$ can be. First, $\eps$ must be small compared with the structural dimensions for the regularized crack to be adequately sharp. Second, $\eps$ must be small enough for the PFSS to adequately represent the material strength surface. The latter constraint arises from the construction of the nucleation driving force $\ce$ proposed by Kumar et al.~\cite{KBFLP20} and subsequently adopted by Chockalingam~(2025)~\cite{chockalingam2025construction}, which uses the same functional form as the prescribed strength surface. The strength calibration ensures that the PFSS matches the material strength surface exactly at $n$ independent strength states, where $n$ is the number of material parameters in the strength surface, but away from these states, the PFSS generally deviates from the material strength surface at finite $\eps$, and the deviation grows with $\eps$. For commonly used analytic strength surfaces, including Drucker--Prager, Mohr--Coulomb, Mogi--Coulomb, and 3D Hoek--Brown, $n=2$, so that exact matching is enforced at only two independent strength states. Consequently, accurately reproducing the full strength surface has required $\eps$ to be commensurate with the Irwin characteristic length $\lch$~\cite{Chockalingam2026EFM}. As we will demonstrate in this work, the PFSS of these analytic surfaces is a poor approximation of the material strength surface already at $\eps=0.5\,\lch$, even when the analytic surface describes the material strength exactly. This limitation is particularly consequential for materials and structures in which $\lch$ is much smaller than the structural dimensions, such as rock masses encountered in tunneling, borehole stability, and hydraulic fracturing, as well as large concrete structures. In such applications, the need to accurately reproduce the material strength surface, rather than to resolve a sufficiently sharp crack, becomes the more restrictive constraint on $\eps$, necessitating extremely fine meshes and substantial computational expense that can be difficult to mitigate even with adaptive mesh refinement~\cite{Nagaraja2019hpFCM,Dahal2026coarsening,Dahal2026localization}. Using a more highly parameterized strength function, rather than conventional analytic forms with few parameters, could improve the accuracy of the PFSS at larger regularization lengths.\\

Both limitations therefore call for the same remedy: a flexible, highly parameterized strength function that is directly learned from data. Machine learning offers a natural and powerful pathway to such a description. Neural networks in particular are highly parameterized function approximators, and machine learning of constitutive and failure behavior has matured considerably~\cite{Fuhg2024review}. Data-driven constitutive models~\cite{KleinLinkaKuhl2022,LinkaKuhl2023CMAME,Tac2022polyconvex} have demonstrated that neural-network functional forms can represent complex multiaxial response while respecting physical constraints such as polyconvexity or thermodynamic consistency~\cite{Linden2023guide}, and physics-augmented data-driven damage and failure models have been formulated for hyperelastic soft materials, including Mullins-type, anisotropic and gradient-enhanced damage~\cite{BuganzaTepole2024preprint,Upadhyay2024MLfracture,Zlatic2024Mullins,AmiriHezaveh2026damage,Wilkinson2026JAXFEM}. None of these, however, addresses the strength surface that governs fracture nucleation.\\

A difficulty that hampers the learning of strength surfaces from experimental data is that strength data are limited. A machine-learned strength function should therefore be constrained by physics, so that it does not produce unphysical predictions under loadings not covered by the data. Recently, star-convexity with respect to the origin has been identified as the central physical requirement on any admissible strength surface~\cite{Chockalingam2026EFM,KumarYavari2026starconvexity}, providing the precise physical constraint that a learned strength surface must respect. In addition, many brittle materials do not fail under certain loadings, such as hydrostatic-compression-dominated ones, up to extremely large stresses, so the strength function must also allow the strength surface to be open along such loading directions.\\

In this work, we introduce a neural-network (NN) strength function for isotropic brittle materials and incorporate it into the strength-incorporated phase-field theory while respecting the physical requirements outlined above. A single NN strength function, with one architecture and training procedure, represents the strength surfaces of a wide range of materials, which removes the need to select and calibrate a different analytic form for each material. It captures the widely used analytic strength surfaces, learns directly from noisy and incomplete experimental strength data, and represents strength behavior that the typical analytic forms do not capture, such as non-convex strength surfaces. The NN strength function is star-convex with respect to the origin by construction, so that the learned surface satisfies this physical requirement for any values of the network parameters, and loading directions along which the material does not fail can be represented as open by choice. Crucially, the NN strength function has far more material parameters than the analytic strength surfaces, which allows a large number of strength states to be calibrated exactly at any $\eps$, so that the PFSS reproduces the material strength surface accurately even at large $\eps$. To make full use of these parameters, we further generalize the calibration solution of Chockalingam~(2025) in two ways. First, the calibration is cast as a least-squares problem, so that any number of calibration strength states can be matched, in a least-squares sense, independently of the parameter dimensionality $n$ and without requiring the chosen states to be independent. Second, openness of the PFSS is enforced, through constraints in a quadratic program, along chosen loading directions in which the material strength surface is open.\\

We will demonstrate the capability of the single NN strength function to accurately capture synthetic strength data generated from the Mohr--Coulomb, Drucker--Prager, Mogi--Coulomb and 3D Hoek--Brown surfaces, as well as experimental strength data for graphite, NR and SBR. The PFSS of the learned strength surfaces remains accurate up to regularization lengths as large as $\eps=10\,\lch$, and even at $\eps=0.5\,\lch$ it is far more accurate than the PFSS of the exact analytic surfaces from which the synthetic data were generated. Finite-element simulations further confirm that the phase-field theory with the learned strength surface captures fracture nucleation under uniform stress, crack growth from large pre-existing cracks, and the transition between strength- and toughness-controlled fracture. By removing the strength-based limitation on the regularization length, the present approach substantially reduces the mesh requirements of phase-field simulations of fracture in large structures, particularly in three dimensions.\\

We note that very recently, Nasir et al.~(2027)~\cite{Nasir2027} incorporated a machine-learned strength surface, represented by an input-convex neural network, into a phase-field model of fracture. Their approach differs from the present one in three respects. First, the driving force is constructed in a different manner: the strength enters through a nucleation term proportional to the strength function that is blended with the classical Griffith-type phase-field driving force by a switching function of the phase field, together with a correction factor that plays a role analogous to that of $\de$ in the present formulation.\footnote{Nasir et al.~\cite{Nasir2027} state that incorporating a new type of strength surface in the theory of Kumar et al.\ requires re-deriving $\ce$, that data-driven strength surfaces require a new formulation, and that an approximate closed-form expression for $\de$ exists only for the Drucker--Prager surface. However, the general expression for $\ce$ of Chockalingam~(2025)~\cite{chockalingam2025construction} applies without re-derivation to any strength function, including the NN strength function of the present work (\Cref{sec:nn_F}), and approximate closed-form expressions for $\de$ have also been provided for the Mohr--Coulomb, Mogi--Coulomb and 3D Hoek--Brown surfaces~\cite{Chockalingam2026EFM}.} Second, the learned surface is constrained to be convex, a requirement stricter than the physically required star-convexity with respect to the origin, which excludes non-convex strength surfaces such as those of the elastomers considered in this work. Third, the regularization lengths considered are at most of the order of $\lch$, so the large regularization lengths that are the focus of the present work are not examined.\\

The remainder of this paper is organized as follows. In \Cref{sec:theory}, we summarize the strength-incorporated phase-field theory of Kumar et al.~\cite{KFLP18,KBFLP20} and the driving force of Chockalingam~(2025), and present its generalized calibration. In \Cref{sec:strength_functions}, we introduce the NN strength function and its training, and summarize the analytic strength surfaces considered. \Cref{sec:results_learning} presents the machine learning of strength surfaces from synthetic and experimental strength data, together with the corresponding PFSS. \Cref{sec:results_fea} presents the finite-element validation of the phase-field theory with the NN strength function, demonstrating strength-based nucleation, toughness-based propagation and the strength--toughness transition in canonical fracture tests, and discusses the computational implications of a large regularization length. Concluding remarks are provided in \Cref{sec:conclusions}.

\section{Theory}\label{sec:theory}

We begin by summarizing the strength-incorporated phase-field theory of Kumar et al.~\cite{KFLP18,KBFLP20} and the general driving-force solution by Chockalingam~(2025)~\cite{chockalingam2025construction} to incorporate arbitrary strength surfaces; we condense it here for completeness and to make the present work self-contained. For detailed derivations, see~\cite{chockalingam2025construction,Chockalingam2026EFM}. A generalization of the driving-force solution presented in~\cite{chockalingam2025construction} is introduced in \Cref{sec:generalized_calib} to account for (i) an arbitrary number of strength calibration locations, and (ii) the enforcement of surface openness in chosen loading states.

\subsection{Kinematics and basic definitions}\label{sec:kinematics}

Consider a structure made of an isotropic elastic brittle material occupying an open bounded domain $\mathrm{\Omega}\subset \mathbb{R}^3$, with boundary $\partial\mathrm{\Omega}$ and outward unit normal $\nten{N}$, in its undeformed and stress-free configuration at time $t=0$. At a later time $t \in (0, T]$, due to an externally applied displacement $\tilde{\bfu}(\bfX, t)$ on a part $\partial\mathrm{\Omega}_\mathcal{D}$ of the boundary and a traction $\tilde{\textbf{t}}(\bfX,t)$ on the complementary part $\partial\mathrm{\Omega}_\mathcal{N}=\partial\mathrm{\Omega}\setminus \partial\mathrm{\Omega}_\mathcal{D}$, the material points described by position vector $\nten{X}$ experience a displacement field $\nten{u}(\nten{X},t)$ and are displaced to the new position $\bfx = \bfX + \bfu(\bfX,t)$. The deformation gradient is
\begin{equation}
\bfF(\bfu)=\bfI+\nabla\bfu,
\end{equation}
with material impenetrability requiring $J \equiv {\rm det}\,\bfF>0$. For the materials considered in this work that are linear elastic, the deformations preceding brittle failure are infinitesimal, and the relevant strain measure is the linearized strain tensor
\begin{equation}
\bfE(\bfu)=\dfrac{1}{2}\!\left(\bfF+\bfF^T\right)-\bfI=\dfrac{1}{2}(\nabla\bfu+\nabla\bfu^T).
\end{equation}
Cracks are described in a regularized fashion via a phase field
\begin{equation}
v=v(\bfX,t)\in[0,1],
\end{equation}
with $v=1$ identifying regions of the sound material and $0\le v<1$ identifying regions undergoing fracture.

\subsection{Constitutive behavior}\label{sec:constitutive}

The mechanical behavior of a brittle material is taken to be fully characterized by three intrinsic properties: \emph{(i)} elasticity, \emph{(ii)} strength, and \emph{(iii)} toughness, defined as follows.

\subsubsection{Elasticity}\label{sec:elasticity}

The elasticity of the material is characterized by a stored-energy function
\begin{equation}
W = W(\bfF)\label{eq:WofF}
\end{equation}
defined for arbitrary deformation gradients, from which the first Piola--Kirchhoff stress follows as $\bfS = \partial W(\bfF)/\partial \bfF$ and the Cauchy stress as $\bm{\sigma} = J^{-1}\bfS\,\bfF^{T}$. Three specific forms of $W$ are employed in this work; \Cref{tab:material_params} records which form is used for which material.

\runin{Linear elasticity} For the rocks and ceramics considered here, the deformations preceding brittle failure are infinitesimal and $W$ reduces to
\begin{equation}
W(\bfE(\bfu)) =\mu \, {\rm tr}\,\bfE^2+\dfrac{\lambda}{2}({\rm tr}\,\bfE)^2,\label{eq:W-mu}
\end{equation}
with $\mu>0$ and $\lambda>-\tfrac{2}{3}\mu$ the Lam\'e constants ($\mu=E/(2(1+\nu))$, $\lambda=E\nu/((1+\nu)(1-2\nu))$, where $E$ and $\nu$ are Young's modulus and Poisson's ratio). The Cauchy and first Piola--Kirchhoff stresses then coincide to leading order and are given by
\begin{equation}
\bm{\sigma} = \pdv{W(\nten{E})}{\nten{E}} = 2\mu\bfE+\lambda({\rm tr}\,\bfE)\bfI.\label{eq:Stress}
\end{equation}

\runin{Yeoh} For natural rubber (\Cref{sec:exp_nr}) the strain energy is  {described using the} incompressible hyperelastic Yeoh form~\cite{Yeoh1990,hamdi2006rubber}
\begin{equation}
W_{\text{Yeoh}} = C_{10}(I_1^c-3) + C_{20}(I_1^c-3)^2 + C_{30}(I_1^c-3)^3,\label{eq:Yeoh}
\end{equation}
where $I_1^c = \lambda_1^2+\lambda_2^2+\lambda_3^2 = {\rm tr}\,(\bfF^T\bfF)$ is the first invariant of the right Cauchy--Green tensor and the principal stretches $\lambda_i$ satisfy the incompressibility constraint $\lambda_1\lambda_2\lambda_3=1$. %

\runin{Ogden} For styrene--butadiene rubber (\Cref{sec:exp_sbr}) the strain energy is  {described using the} incompressible hyperelastic two-term Ogden form~\cite{Ogden1972,hamdi2006rubber}
\begin{equation}
W_{\text{Ogden}} = \sum_{k=1}^{2}\frac{\mu_k}{\alpha_k}\!\left(\lambda_1^{\alpha_k}+\lambda_2^{\alpha_k}+\lambda_3^{\alpha_k}-3\right),\label{eq:Ogden}
\end{equation}
again with $\lambda_1\lambda_2\lambda_3=1$.

\runin{Stress response of the incompressible hyperelastic forms} For both the Yeoh and Ogden strain energies, the principal Cauchy stresses are
\begin{equation}
\sigma_i=\lambda_i\pdv{W}{\lambda_i}-p,\qquad i=1,2,3\ \text{(no sum)},\label{eq:Cauchy_incomp}
\end{equation}
which for~\cref{eq:Yeoh,eq:Ogden} read
\begin{equation}
\text{Yeoh:}\quad \sigma_i=2\,\Lambda\,\lambda_i^2-p,\quad \Lambda=C_{10}+2C_{20}(I_1^c-3)+3C_{30}(I_1^c-3)^2;\qquad
\text{Ogden:}\quad \sigma_i=\sum_{k=1}^{2}\mu_k\lambda_i^{\alpha_k}-p.\label{eq:sigma_YO}
\end{equation}
Here $p$ is a Lagrange multiplier associated with the incompressibility constraint $\lambda_1\lambda_2\lambda_3=1$, determined through the governing equations and boundary conditions. Taking differences of~\cref{eq:Cauchy_incomp} eliminates $p$:
\begin{equation}
\sigma_i-\sigma_3=\lambda_i\pdv{W}{\lambda_i}-\lambda_3\pdv{W}{\lambda_3},\qquad i=1,2,\qquad \lambda_3=\frac{1}{\lambda_1\lambda_2}.\label{eq:Cauchy_diff}
\end{equation}
For a given stress state, the two equations in~\cref{eq:Cauchy_diff} can be solved for $(\lambda_1,\lambda_2)$, which inverts the stress--deformation relation. For plane stress ($\sigma_3=0$),~\cref{eq:Cauchy_diff} directly gives the in-plane stress--stretch relations.

\runin{Stress invariants} Denoting the principal stresses by $\sigma_1,\sigma_2,\sigma_3$ with $\sigmax = \max(\sigma_1,\sigma_2,\sigma_3)$ and $\sigmin = \min(\sigma_1,\sigma_2,\sigma_3)$, we introduce the invariants
\begin{equation}
I_1 = \sigma_1+\sigma_2+\sigma_3,\quad J_2 = \tfrac{1}{6}\!\left((\sigma_1-\sigma_2)^2+(\sigma_2-\sigma_3)^2+(\sigma_3-\sigma_1)^2\right),\quad \sigmTwo = \tfrac{1}{2}(\sigmax+\sigmin),\label{eq:Invariants}
\end{equation}
together with the determinant of the deviatoric stress $J_3 = \det(\devop(\bm\sigma))${, where $\devop(\bm\sigma)=\bm\sigma-\tfrac{1}{3}({\rm tr}\,\bm\sigma)\bfI$ denotes the deviatoric part of $\bm\sigma$,} and the scaled triaxiality\footnote{The stress triaxiality is conventionally defined as the ratio of the mean stress $\sigma_\mathrm{m}=I_1/3$ to the von Mises equivalent stress $\sigma_\mathrm{vM}=\sqrt{3J_2}$, i.e.\ $\sigma_\mathrm{m}/\sigma_\mathrm{vM}=I_1/(3\sqrt{3J_2})$. Here we use the scaled triaxiality $\Itriax=\sqrt{3}\,\sigma_\mathrm{m}/\sigma_\mathrm{vM}=I_1/(3\sqrt{J_2})$ throughout.} and Lode-angle invariants
\begin{equation}
\Itriax = \frac{I_1}{3\sqrt{J_2}}, \qquad
\Lode = \frac{1}{3}\arcsin\!\left(\frac{3\sqrt{3}}{2}\,\frac{J_3}{J_2^{3/2}}\right) \in \!\left[-\tfrac{\pi}{6},\tfrac{\pi}{6}\right],\label{eq:eta_theta}
\end{equation}
which serve as inputs to the NN strength function introduced in \Cref{sec:nn_F}. 

\runin{Strain energy written in terms of stress} Phase-field calibration equations in \Cref{sec:ce} require the stored energy evaluated as a function of the stress state,
\begin{equation}
W(\bm{\sigma}) \equiv W\!\left(\bfF(\bm{\sigma})\right),\label{eq:Wofsigma_general}
\end{equation}
obtained by inverting the stress--deformation relation at the given stress. For linear elasticity this inversion is explicit and yields the familiar complementary form
\begin{equation}
W(\bm{\sigma}) = \frac{1}{2}\!\left(\frac{J_2}{\mu}+\frac{I_1^2}{9 K}\right),\qquad K=\lambda+\tfrac{2\mu}{3},\label{eq:Wofsigma}
\end{equation}
where $K$ is the bulk modulus. For the Yeoh and Ogden forms no closed-form inversion exists and $W(\bm{\sigma})$ is evaluated numerically{, by solving~\cref{eq:Cauchy_diff} for the deformation field from the stress state and then evaluating the strain energy.} 

\subsubsection{Strength}\label{sec:strength}

The strength of the material controls fracture nucleation in large specimens of homogeneous brittle materials subjected to a uniform state of stress. When such a specimen is subjected to monotonically increasing, spatially uniform stress, a crack will form at an indeterminate location at a critical value of the applied stress. The collection of all such critical stresses defines a surface in stress space, the material's \emph{strength surface}, which is taken here as a material property (modulo stochasticity)~\cite{KBFLP20}. We assume it can be written in the general normalized form
\begin{equation}
\mathcal{F} \equiv F(\bm{\sigma};\myvect{\beta}) = g(\bm{\sigma};\myvect{\beta}) - 1 = 0,\label{eq:SSurf-0}
\end{equation}
where $\myvect{\beta}$ is an $n$-dimensional array of material constants and $g(\bm{0};\myvect{\beta})=0$ so that $F(\bm{0};\myvect{\beta}) = -1$. For any stress state $\bm{\sigma}$ before attainment of strength, $F(\bm{\sigma}) < 0$; the set of such states is the \emph{elastic domain} of the material, while $F(\bm{\sigma}) > 0$ is a violation of strength. The uniaxial tensile and compressive strengths $\sts>0$ and $\scs>0$ are the critical stresses on this surface under uniaxial tension and uniaxial compression, respectively, defined by
\begin{equation}
F(\sts\,\bfe_1\otimes\bfe_1;\myvect{\beta})=0,\qquad F(-\scs\,\bfe_1\otimes\bfe_1;\myvect{\beta})=0.\label{eq:sts_scs}
\end{equation}

\runin{Star-convexity with respect to the origin}\label{par:starconvex}
A central physical requirement on any admissible strength surface for a brittle material is \emph{star-convexity with respect to the origin}~\cite{Chockalingam2026EFM,KumarYavari2026starconvexity}: along any radial direction in stress space, the strength surface is crossed at most once (exactly once along directions in which the surface is closed, and never along open directions). Formally, the surface $F(\bm{\sigma})=0$ is star-convex with respect to the origin if
\begin{equation}
F(\kappa \bm{\sigma}) < 0 \quad \text{for all } \kappa \in [0,1)\;\; \text{whenever}\;\; F(\bm{\sigma})=0,\label{eq:starconvexity}
\end{equation}
i.e.\ the line segment from the origin to any point on the strength surface lies entirely in the elastic domain. Equivalently, for proportional loading along any loading direction $\hat{\bm{\sigma}}$ with $\lVert\hat{\bm{\sigma}}\rVert=1$, such that $\bm{\sigma} = \lVert\bm{\sigma}\rVert\,\hat{\bm{\sigma}}$, where $\lVert\bm{\sigma}\rVert=\sqrt{\bm{\sigma}:\bm{\sigma}}$, $F$ changes sign in $\lVert\bm{\sigma}\rVert$ at most once. This is the mathematical translation of the physical statement that a material cannot fail and then return to an elastic state at a larger stress magnitude along the same loading direction. \textit{Star-convexity is therefore a non-negotiable requirement on any data-driven or analytic strength surface; the neural-network construction of \Cref{sec:nn_F} enforces it by design.}

\begin{remark}\label{rem:notconvex}{\rm The strength surface $\mathcal{F}$ need not be convex, unlike the yield surfaces in associative plasticity, which are often represented in a similar form. Non-convexity is in fact experimentally observed---most famously in brittle elastomers, where the material's weak behavior under hydrostatic tension results in a non-convex strength surface~\cite{KKLP24}. Additional discussions on convexity of strength and yield surfaces can be found in~\cite{bigoni2004yield,cai2021new}.}\end{remark}

\begin{remark}\label{rem:homog}{\rm Any strength function of the form in~\cref{eq:SSurf-0} whose $g$ is positively homogeneous of degree one in $\bm\sigma$, i.e.\ $g(\kappa\bm\sigma;\myvect{\beta})=\kappa\,g(\bm\sigma;\myvect{\beta})$ for all $\kappa\ge0$, is automatically star-convex with respect to the origin. Along any ray $\bm\sigma=\lVert\bm\sigma\rVert\,\hat{\bm\sigma}$,
\begin{equation*}
F(\bm\sigma;\myvect{\beta})=\lVert\bm\sigma\rVert\,g(\hat{\bm\sigma};\myvect{\beta})-1
\end{equation*}
is linear in $\lVert\bm\sigma\rVert$ and starts from $F=-1$ at the origin. Along loading directions $\hat{\bm\sigma}$ with $g(\hat{\bm\sigma};\myvect{\beta})>0$, $F$ crosses zero exactly once, at $\lVert\bm\sigma\rVert=1/g(\hat{\bm\sigma};\myvect{\beta})$, where strength failure occurs; along directions with $g(\hat{\bm\sigma};\myvect{\beta})\le0$, $F\le-1$ for all $\lVert\bm\sigma\rVert$, so that no strength failure occurs and the surface is open in that direction. In either case the surface is crossed at most once, as required by~\cref{eq:starconvexity}. Several widely used analytic strength functions are homogeneous of degree one, including the Rankine, Tresca, von Mises, Mohr--Coulomb, Drucker--Prager and Mogi--Coulomb functions; for Drucker--Prager, for instance, the directions with $g\le0$ are precisely its open directions under compressive loading~\cite{chockalingam2025construction}. Strength functions that are not homogeneous of degree one, such as the Hoek--Brown and 3D Hoek--Brown functions~\cite{Chockalingam2026EFM}, are not covered by this argument, but their star-convexity can be established separately~\cite{KumarYavari2026starconvexity}.}\end{remark}

\textit{In this work, the NN strength function is constructed to be positively homogeneous of degree one in $\bm\sigma$ (\Cref{sec:nn_F}), which guarantees its star-convexity by Remark~\ref{rem:homog}.}

\subsubsection{Fracture toughness and Irwin characteristic length}\label{sec:toughness}

The critical energy release rate (fracture toughness), denoted~$G_c$, controls fracture from a large pre-existing crack~$\Gamma$ through the Griffith criticality condition~\cite{Griffith21}
\begin{equation}
G = -\dfrac{\partial{\mathcal{W}}}{\partial \Gamma}=G_c,\label{eq:Griffith}
\end{equation}
where $\mathcal{W}$ is the potential energy and $G$ is the energy release rate. A pre-existing crack advances when $G$ reaches $G_c$. The notion of a `large' crack means that the crack and the body dimensions are much larger than the characteristic process-zone size, measured by the Irwin characteristic length as %
\begin{equation}
\lch = \frac{3\,G_c}{16\,W_{\mathtt{ts}}},\qquad W_{\mathtt{ts}} \equiv W\!\left(\sts\,\bfe_1\otimes\bfe_1\right).\label{eq:lch}
\end{equation}
Here $W_{\mathtt{ts}}$ is the stored energy in~\cref{eq:Wofsigma_general} evaluated at the state of uniaxial tension at the tensile strength. For a linear-elastic material $W_{\mathtt{ts}}=\sts^2/(2E)$, and~\cref{eq:lch} reduces to the familiar Irwin form $\lch = 3EG_c/(8\sts^2)$. %

\subsection{Phase-field governing equations}\label{sec:phasefield}

Under general loadings and non-uniform stress states, an interplay between strength and fracture toughness governs fracture behavior. The phase-field theory of Kumar et al.~\cite{KFLP18,KBFLP20} incorporates strength via a driving force $\ce(\nten{X},t)$ introduced into the Euler--Lagrange equations of the classical phase-field theory~\cite{Bourdin00,Bourdin08}, which already encode the Griffith criticality condition. The displacement field $\bfu_k(\bfX)=\bfu(\bfX,t_k)$ and phase field $v_k(\bfX)=v(\bfX,t_k)$ at material point $\bfX\in\overline{\Omega}$ and discrete time $t_k\in\{0=t_0,t_1,\ldots,t_M=T\}$ are determined by
\begin{subequations}\label{eq:BVP-u}
\begin{empheq}[left=\empheqlbrace]{align}
&{\rm Div}\!\left[v_{k}^2\dfrac{\partial W}{\partial \bfF}(\bfF(\bfu_{k}))\right]={\bf0}, \quad \text{if}~\bfX \in \Omega, \label{eq:BVP-u-a}\\[3pt]
&\bfu_{k}=\tilde{\bfu}(\bfX,t_{k}), \quad \text{if}~\bfX \in \partial\mathrm{\Omega}_\mathcal{D}, \label{eq:BVP-u-b}\\[3pt]
&\left[v_{k}^2\dfrac{\partial W}{\partial \bfF}(\bfF(\bfu_{k}))\right]\bfN=\tilde{\textbf{t}}(\bfX,t_{k}), \quad \text{if}~\bfX \in \partial\mathrm{\Omega}_\mathcal{N}, \label{eq:BVP-u-c}
\end{empheq}
\end{subequations}
and
\begin{subequations}\label{eq:BVP-v}
\begin{empheq}[left=\empheqlbrace]{align}
&\tfrac{3}{4}\varepsilon\,\de\,G_c \triangle v_{k}
   = 2 v_{k} W(\bfF(\bfu_{k})) - \ce(\nten{X},t_k)
   - \tfrac{3\,\de\,G_c}{8\varepsilon},\quad
   \text{if } v_{k}(\bfX)< v_{k-1}(\bfX),\quad
   \bfX\in\Omega, \label{eq:BVP-v-main}\\[4pt]
&\tfrac{3}{4}\varepsilon\,\de\,G_c \triangle v_{k}
   \ge 2 v_{k} W(\bfF(\bfu_{k})) - \ce(\nten{X},t_k)
   - \tfrac{3\,\de\,G_c}{8\varepsilon},~
   \text{if } v_{k}(\bfX)=1 \text{ or } v_{k}(\bfX)= v_{k-1}(\bfX)>0,\,
   \bfX\in\Omega, \label{eq:BVP-v-secondary}\\[4pt]
&v_{k}(\bfX)=0,\quad
   \text{if } v_{k-1}(\bfX)=0,\quad
   \bfX\in\Omega, \label{eq:BVP-v-zero}\\[4pt]
&\nabla v_{k}\cdot\bfN=0,\quad
   \bfX\in \partial\Omega, \label{eq:BVP-v-bc}
\end{empheq}
\end{subequations}
with $\triangle(\cdot) \equiv {\rm Div}(\nabla(\cdot))$ and $\eps > 0$ the regularization length---the length scale over which a sharp crack is smeared by the phase field.  
The parameter $\de > 0$ is a scalar correction factor that helps preserve correct fracture toughness in Griffith physics, as explained later in this section. Setting $\de=1$ and $\ce=0$ recovers the classical phase-field theory~\cite{Bourdin00,Bourdin08}, which captures Griffith propagation but not strength-controlled nucleation~\cite{lopez2025classical,kamarei2025nucleation}. The degraded/true first Piola--Kirchhoff stress and Cauchy stress are given by
\begin{equation}
\bfS = v^2\,\pdv{W}{\bfF}, \quad \bm{\sigma}= v^2 J^{-1}\pdv{W}{\bfF}\,\bfF^{T}. \label{eq:truestresses}
\end{equation}

\runin{Regularization length} The regularization length $\eps$ is regarded here purely as a numerical regularization parameter, with no physical meaning ascribed to it such as an identification with the fracture process-zone size. The mesh size $h$ for accurate simulations is limited by the regularization length $\eps$, as typically $h<\eps$: the finite-element discretization of the regularized crack overestimates the energy dissipated per unit crack area, so that the numerical fracture toughness exceeds $G_c$ by an amount that grows with $h/\eps$~\cite{Bourdin00,Bourdin08}. The mesh size in fracturing regions must therefore be a small fraction of $\eps$ for the toughness-controlled (Griffith) response to be captured accurately; in this work, $h=\eps/5$ in fracturing regions (\Cref{sec:fea_cost}). Thus the number of nodes in the simulation and simulation time are greatly dependent on how small the regularization length needs to be. There are two factors that limit the size of the regularization length $\eps$: (i) in problems involving cracks and notches, it needs to be small compared with the geometric length scales of the problem, such as the structural dimensions and the sizes of cracks and notches, for the regularized crack representation to be adequately sharp, and (ii) for problems where the strength of the material plays a role, $\eps$ also needs to be small enough to get an adequate representation of the strength surface of the material, as described later in~\Cref{sec:pfss,sec:results_learning}. For simulations involving large structural dimensions, the latter becomes the critical limiting factor on how large $\eps$ can be. This drastically increases the simulation times for large structures. \textit{In this manuscript, we will show that a key advantage of the use of NN strength functions is that they allow for accurate strength representation at large $\eps$, compared to analytic strength functions, which alleviates the strength-based limitation on the regularization length.}

\subsubsection{Driving force $\ce$ and exact strength calibration}\label{sec:ce}

A roadmap for constructing $\ce$ was presented in Kumar et al.~(2020)~\cite{KBFLP20}, where explicit expressions were provided for the Drucker--Prager strength surface. Following this roadmap, Chockalingam~(2025)~\cite{chockalingam2025construction} derived the following expression for $\ce$ for a general strength surface of the form in~\cref{eq:SSurf-0},
\begin{equation}
\ce = -\omega_\eps\,g\!\left(\bm{\sigma};\myvect{\beta}+\Dbe\right),\qquad
\omega_\eps = \frac{3}{8}\,\frac{\de\,G_c}{\eps},\label{eq:general_ce}
\end{equation}
where $\bm\sigma$ is the degraded true stress from~\cref{eq:truestresses}. This driving force $\ce$ allows for the strength surface $\mathcal{F}$ of the material to be recovered exactly in the limit of vanishing regularization length $\eps\searrow0$. %
The correction parameters $\Dbe$ are chosen so that $n$ independent calibration strength states $\bm{\sigma}_{\texttt{s}i}$, $i=1,\ldots,n$, chosen on the material strength surface (see Remark~\ref{rem:beta_independence} for the definition of independence), are also recovered exactly by the phase-field theory at any finite value of $\eps$ (recall that $n$ is the dimensionality of the material parameters $\myvect{\beta}$ in~\cref{eq:SSurf-0}). Specifically,
\begin{equation}
2\,\frac{W(\bm{\sigma}_{\texttt{s}i})}{\omega_\eps} + F\!\left(\bm{\sigma}_{\texttt{s}i};\,\myvect{\beta}+\Dbe\right) = 0, \qquad i=1,\ldots,n. \label{eq:surfacematching}
\end{equation}
When $F$ is linear in $\myvect{\beta}$, $F(\bm{\sigma}_{\texttt{s}i};\myvect{\beta}+\Dbe)=F(\bm{\sigma}_{\texttt{s}i};\myvect{\beta})+\sum_j A_{ij}\,\Delta\beta^\eps_j$, and~\cref{eq:surfacematching} becomes the linear system $\mymatrix{A}\,\Dbe=\myvect{b}$ with
\begin{equation}
A_{ij} = \frac{\partial F(\bm{\sigma}_{\texttt{s}i};\myvect{\beta})}{\partial \beta_j}, \qquad
b_i = -F(\bm{\sigma}_{\texttt{s}i};\myvect{\beta})-\frac{2\,W(\bm{\sigma}_{\texttt{s}i})}{\omega_\eps}, \qquad i,j=1,\ldots,n,\label{eq:Ab_def}
\end{equation}
whose closed-form solution is~\cite{chockalingam2025construction}
\begin{equation}
\Dbe = \mymatrix{A}^{-1}\myvect{b}.\label{eq:deltabetaeps_sol}
\end{equation}
For calibration states lying exactly on the material strength surface, the term $F(\bm{\sigma}_{\texttt{s}i};\myvect{\beta})$ in~\cref{eq:Ab_def} vanishes and thus $b_i$ reduces to $-2W(\bm{\sigma}_{\texttt{s}i})/\omega_\eps$ as in~\cite{chockalingam2025construction}; here we retain the  term because, in some of the neural-network cases considered later,  the training data are chosen as calibration strength states (instead of the strength values on the learned surface), where the learned strength function vanishes only approximately.
As $\eps\searrow 0$, $\omega_\eps\nearrow\infty$ and hence $\Dbe\searrow\myvect{0}$ (assuming $F(\bm{\sigma}_{\texttt{s}i};\myvect{\beta})\approx0$). %

\begin{remark}\label{rem:beta_independence}{\rm
{A set of chosen calibration strength states $\bm{\sigma}_{\texttt{s}i}$ is said to be \emph{independent} when the corresponding rows of the matrix $\mymatrix{A}$ in~\cref{eq:Ab_def} are linearly independent. For $m$ calibration strength states to be independent, we necessarily require $m\le n$, and thus the maximum number of strength states that can be independent is $n$ (the dimensionality of the material parameters $\myvect{\beta}$). If the total number of calibration strength states is chosen to be equal to $n$, as in~\cref{eq:surfacematching}, $\mymatrix{A}$ is square, and all the chosen strength states being independent is equivalent to $\mymatrix{A}$ being invertible~\cite{chockalingam2025construction}. In such a case, all $n$ independent strength states can be calibrated exactly.} If the chosen calibration strength states are not independent, the inversion in~\cref{eq:deltabetaeps_sol} becomes ill-defined---this can become an issue for strength surfaces with high parameter dimensionality $n$ where carefully choosing independent strength states might become difficult. This is addressed by the generalization of the solution for $\Dbe$ introduced later in~\Cref{sec:generalized_calib} which allows for the calibration strength states to be recovered in a least-squares sense.}\end{remark}

\begin{remark}\label{rem:beta_linearity}{\rm
The closed-form solution {in}~\cref{eq:deltabetaeps_sol} is explicit \emph{only} when $F$ is linear in $\myvect{\beta}$. The analytic surfaces in \Cref{sec:analytic_F} are all linear in their $\myvect{\beta}$ by construction. For the NN strength function $\FNN$ introduced in \Cref{sec:nn_F}, $\FNN$ is generally nonlinear in its parameters, but only the linear output weights are used as $\myvect{\beta}$ in the calibration here.}\end{remark}

In the presence of large cracks (toughness-dominated regime), the driving force $\ce$ modifies the equations of the classical phase-field theory, and there is no a priori guarantee that the resulting theory still follows Griffith's criticality condition in~\cref{eq:Griffith}. Numerical studies show, however, that it does, but with an effective critical energy release rate that differs from $G_c$. The scalar $\de$ corrects this effective value to $G_c$, and is calibrated once from any boundary-value problem for which nucleation from a large pre-existing crack is known exactly from~\cref{eq:Griffith}. The evidence for the Griffith-type behavior is then that the same $\de$ yields the correct response in other boundary-value problems, with different geometries, crack lengths and loadings~\cite{KBFLP20,KRLP22,KKLP24,LK24,KDK2025Comparison,KAMAREI2026118449,Chockalingam2026EFM}; we verify this for the NN strength function in \Cref{sec:results_fea}. Approximate analytic forms of $\de$ as a function of mesh size and $\eps/\lch$ have also been presented previously for the Drucker--Prager, Mohr--Coulomb, 3D~Hoek--Brown, and Mogi--Coulomb surfaces~\cite{KKLP24,Chockalingam2026EFM}.

\subsubsection{Phase-field strength surface (PFSS)}\label{sec:pfss}

The strength surface $\mathcal{F}^\eps$ generated by the strength-incorporated phase-field theory for the driving force presented in the earlier section is given by the following equation~\cite{KBFLP20,chockalingam2025construction}:
\begin{equation}
\mathcal{F}^\eps \equiv {F}^\eps = 0, \qquad {F}^\eps \equiv \frac{2 W(\bm{\sigma})}{\omega_\eps} + F\!\left(\bm{\sigma};\,\myvect{\beta}+\Dbe\right). \label{eq:PFSS_general}
\end{equation}
This is the \emph{phase-field strength surface} (PFSS). It is obtained directly by setting $v_k=1$ in~\cref{eq:BVP-v-main}; numerical evidence~\cite{KBFLP20,KLP20} shows that for sufficiently large specimens the initially uniform phase field loses stability and localizes into a crack at the stresses defined by~\cref{eq:PFSS_general}. As $\eps\searrow 0$ (and hence $\omega_\eps\nearrow\infty,~\Dbe\searrow \myvect{0}$), $\mathcal{F}^\eps\to\mathcal{F}$, i.e.\ the PFSS collapses to the material strength surface. At finite $\eps$, $\mathcal{F}^\eps$ deviates from $\mathcal{F}$ everywhere except at the $n$ calibration locations $\bm{\sigma}_{\texttt{s}i}$, where matching is exact by construction {in}~\cref{eq:surfacematching}.\\ %

A key concern for phase-field modeling is the quality of the PFSS approximation of the material strength surface at finite~$\eps$. The deviation is controlled by~$\omega_\eps^{-1}$ (which increases with regularization length $\eps$) and by the parameterization $n$ of $\myvect{\beta}$ (which sets how many calibration strength locations can be matched exactly). For most analytic strength surfaces in widespread use, $n$ is small ($n=2$ for M--C, Drucker--Prager (D--P), Mg--C, 3D H--B)---hence, at larger $\eps$, the analytic-form PFSS is a poor approximation of the material strength surface everywhere away from the few calibration locations. Later, we will show that the high parameter dimensionality of the NN strength function allows it to capture the material strength surface accurately across regularization lengths.

\runin{Compression correction} For low-parameter analytic surfaces, a phenomenological correction has been employed in prior work~\cite{KRLP22,Chockalingam2026EFM} that switches off the strain-energy contribution in~\cref{eq:PFSS_general} in compressive stress states when $I_1<0$ (assuming fracture is not expected under such compressive stress states), by replacing $W(\bm{\sigma})\mapsto W(\bm{\sigma})\,f_\texttt{cc}(I_1)$ with $f_\texttt{cc}(I_1)=1$ for $I_1\ge 0$ and $0$ for $I_1<0$. This improves the PFSS in compressive regions for analytic surfaces but introduces a discontinuity at $I_1=0$. We note here, and demonstrate in \Cref{sec:results_learning}, that the high parameterization of the NN strength function makes this correction unnecessary: the regular PFSS in~\cref{eq:PFSS_general} already matches the material strength surface accurately across the entire stress space (including in compression) up to large~$\eps/\lch$.

\runin{Star-convexity of the PFSS with respect to the origin}\label{par:pfss_starconvex}
Star-convexity of the PFSS defined in~\cref{eq:PFSS_general} with respect to the origin is not necessarily guaranteed from that of the material strength surface. However, it can be shown to hold necessarily for linear-elastic materials whose strength function is positively homogeneous of degree one. As shown in Remark~\ref{rem:homog}, for such materials the function $g$ is linear in $\lVert\bm\sigma\rVert$. The strain energy for linear-elastic materials satisfies $W(\bm\sigma)=\lVert\bm\sigma\rVert^2\,W(\hat{\bm\sigma})$ with $W(\hat{\bm\sigma})>0$. Along every ray $\bm\sigma=\lVert\bm\sigma\rVert\,\hat{\bm\sigma}$ the PFSS function in~\cref{eq:PFSS_general} therefore reads
\begin{equation*}
{F}^\eps(\bm\sigma)=a\lVert\bm\sigma\rVert^2+b\lVert\bm\sigma\rVert-1,\qquad a=\frac{2W(\hat{\bm\sigma})}{\omega_\eps}>0,\qquad b=g(\hat{\bm\sigma};\myvect{\beta}+\Dbe)\ \text{of either sign},
\end{equation*}
a quadratic with exactly one positive root, so that the PFSS is star-convex. The NN strength function introduced in \Cref{sec:nn_F} and the Mohr--Coulomb, Mogi--Coulomb and Drucker--Prager strength functions {of} \Cref{sec:analytic_F} are all positively homogeneous of degree one; their PFSS for linear-elastic materials are therefore star-convex by construction. The same holds with the compression correction, which removes the quadratic term for $I_1<0$ and leaves at most one crossing. No such guarantee holds for strength functions that are not homogeneous of degree one, such as 3D Hoek--Brown (see \Cref{sec:hb}), nor for hyperelastic materials, for which $W$ is not quadratic in $\lVert\bm\sigma\rVert$. {Nevertheless, for practical numerical purposes a loss of star-convexity of the PFSS is typically not an issue, as long as $F^\eps$ does not return to negative values (an elastic state) close to its innermost, i.e.\ most conservative,  surface that satisfies $F^\eps(\bm\sigma)=0$. For all strength surfaces plotted in this work, we verified that $F^\eps<0$ for all stress states inside the plotted failure envelopes.}

\subsubsection{Generalized calibration: least-squares and QP-constrained openness}\label{sec:generalized_calib}

The calibration solution in~\cref{eq:deltabetaeps_sol} from Chockalingam~(2025)~\cite{chockalingam2025construction} assumes that exactly $n$ independent strength locations on the material strength surface are selected for matching on the PFSS. In practice, two situations arise that motivate a generalization: (i) the number $m$ of desired calibration locations may differ from $n$ (the dimensionality of material parameters $\myvect{\beta}$) or the chosen strength states may not be independent (see Remark~\ref{rem:beta_independence}); and (ii) the material strength surface may be \emph{open} in certain loading directions (e.g.\ graphite, titania, and elastomers under biaxial compression), meaning the material does not strength fail under those loadings, which we might want the PFSS to respect so that it does not spuriously close in those directions. We address both within a single least-squares-plus-quadratic-program framework.

\runin{Least-squares calibration} Let $\{\bm{\sigma}_{\texttt{s}i}\}_{i=1}^m$ be the chosen calibration stress states on $\mathcal{F}$, with $m$ now not necessarily equal to $n$. The matrix $\mymatrix{A}\in\mathbb{R}^{m\times n}$ and the vector $\myvect{b}\in\mathbb{R}^m$ are defined exactly as in~\cref{eq:Ab_def}, with the row index now running over $i=1,\ldots,m$. The matching condition {in}~\cref{eq:surfacematching} again reads $\mymatrix{A}\,\Dbe = \myvect{b}$, but this system need not admit an exact or a unique solution, and we therefore take
\begin{equation}
\Dbe \;=\; \argmin_{\myvect{x}\in\mathbb{R}^n}\;\lVert\mymatrix{A}\myvect{x} - \myvect{b}\rVert^2,\label{eq:lsq}
\end{equation}
whose solution is $\Dbe = \mymatrix{A}^{+}\myvect{b}$, where $\mymatrix{A}^{+}$ is the Moore--Penrose pseudoinverse of $\mymatrix{A}$. The pseudoinverse exists and is unique for any $\mymatrix{A}$, so~\cref{eq:lsq} always admits a solution, and $\mymatrix{A}^{+}\myvect{b}$ is the minimizer of smallest norm in~\cref{eq:lsq}.\footnote{{The minimizer is unique when $\rank\mymatrix{A}=n$, and the residual vanishes when $\myvect{b}$ lies in the range of $\mymatrix{A}$; in particular, when the $m$ calibration strength states are independent (Remark~\ref{rem:beta_independence}), i.e.\ $\rank\mymatrix{A}=m\le n$, $\myvect{b}$ always lies in the range of $\mymatrix{A}$ and all calibration strength states are matched exactly. When $\rank\mymatrix{A}<n$ there are infinitely many minimizers; here we select the one with the minimum norm $\lVert\Dbe\rVert$.}} In general the residual in~\cref{eq:lsq} is nonzero, in which case the calibration locations are matched in a least-squares sense rather than exactly. When $\mymatrix{A}$ is invertible, $\mymatrix{A}^{+} = \mymatrix{A}^{-1}$ and~\cref{eq:lsq} reduces to~\cref{eq:deltabetaeps_sol}; the present generalization is therefore strictly broader than the exact solution in Chockalingam~(2025). As $\eps\searrow 0$, $\myvect{b}\searrow\myvect{0}$ in~\cref{eq:Ab_def} and hence $\Dbe\searrow \myvect{0}$ irrespective of the rank of $\mymatrix{A}$, so the limiting consistency of the phase-field theory as defined in~\cite{chockalingam2025construction} is preserved. In this paper, we numerically solve~\cref{eq:lsq} using the singular-value-decomposition-based least-squares routine {\texttt{numpy.linalg.lstsq}} in NumPy~\cite{numpy2020}, which returns the minimum-norm solution.

\runin{Openness via quadratic programming} When the material strength surface is open along certain loading directions $\hat{\bm{\sigma}}_{\texttt{f}j}$, $j=1,\ldots,q$, we additionally require that the PFSS does not predict failure along those directions up to large stress magnitudes, i.e.\ $F^\eps(\bm{\sigma}_{\texttt{f}j})\le 0$, where $\bm{\sigma}_{\texttt{f}j} =  \lVert\bm{\sigma}_{\texttt{f}j}\rVert \hat{\bm{\sigma}}_{\texttt{f}j}$ is a fixed far-field probe stress along the open direction at a large chosen stress magnitude  $\lVert\bm{\sigma}_{\texttt{f}j}\rVert$. While we cannot always enforce true openness of the PFSS along a loading direction, we can effectively force it open for practical purposes by ensuring it does not close before a sufficiently large stress magnitude. For $F$ that is linear in $\myvect{\beta}$, the openness constraint reads 
\begin{equation}
\widetilde{\mymatrix{A}}\,\Dbe + \widetilde{\myvect{b}} \le \myvect{0}
\quad \text{with}\quad 
\widetilde{A}_{ij} = \frac{\partial F(\bm{\sigma}_{\texttt{f}i};\myvect{\beta})}{\partial \beta_j}, \quad
\widetilde{b}_i = \frac{2\,W(\bm{\sigma}_{\texttt{f}i})}{\omega_\eps} + F(\bm{\sigma}_{\texttt{f}i};\myvect{\beta}),\quad i=1,\ldots,\text{{$q$}}.\label{eq:Ab_qp}
\end{equation}
{Here $\widetilde{\mymatrix{A}}\in\mathbb{R}^{q\times n}$ and $\widetilde{\myvect{b}}\in\mathbb{R}^{q}$: each of the $q$ open directions contributes one row, and the inequality in~\cref{eq:Ab_qp} holds componentwise.} Combining with~\cref{eq:lsq}, the corrected calibration is the solution of the convex quadratic program (QP)
\begin{equation}
\Dbe \;=\; \argmin_{\myvect{x}\in\mathbb{R}^n}\;\tfrac{1}{2}\lVert\mymatrix{A}\myvect{x} - \myvect{b}\rVert^2 \quad
\text{subject to}\quad \widetilde{\mymatrix{A}}\,\myvect{x} + \widetilde{\myvect{b}} \;\le\;\myvect{0}.\label{eq:qp}
\end{equation}
 When {$q=0$} (no openness constraints), \cref{eq:qp} reduces to~\cref{eq:lsq}. When the openness constraints are active, exactness at the calibration strength states may be traded for openness. In this manuscript, we numerically solve~\cref{eq:qp} using the \texttt{SLSQP} solver in SciPy~\cite{scipy2020}, which implements the sequential quadratic programming algorithm of Kraft~\cite{Kraft1988}. Together,~\cref{eq:lsq,eq:qp} are the calibration machinery used throughout the rest of this work; in particular, all open surfaces are calibrated with the QP variant, and all closed surfaces with the plain least-squares solver.\\

 This completes the theory development. In the next section we focus on strength functions, where we construct a robust NN strength function and summarize different popular analytic strength functions.
 
\section{Strength functions}\label{sec:strength_functions}

In this section we discuss the different strength functions used in this manuscript. First, in \Cref{sec:nn_F}, we introduce the NN strength function $\FNN$ that is used to fit all the synthetic and experimental data in this manuscript and is a principal contribution of the paper. We then introduce four widely used analytic strength functions in \Cref{sec:analytic_F} that span diverse mathematical and physical characteristics; these strength functions are later used to generate synthetic data in \Cref{sec:syn_results} that are captured using the NN strength function. For the analytic strength functions, we directly record their mathematical forms and the parameter vector $\myvect{\beta}$ in terms of strengths; detailed derivations and physical interpretations can be found in~\cite{chockalingam2025construction,Chockalingam2026EFM} and references therein.

\subsection{Neural-network strength function and training}\label{sec:nn_F}

\begin{figure}[h]
  \centering
  \includegraphics[width=\textwidth]{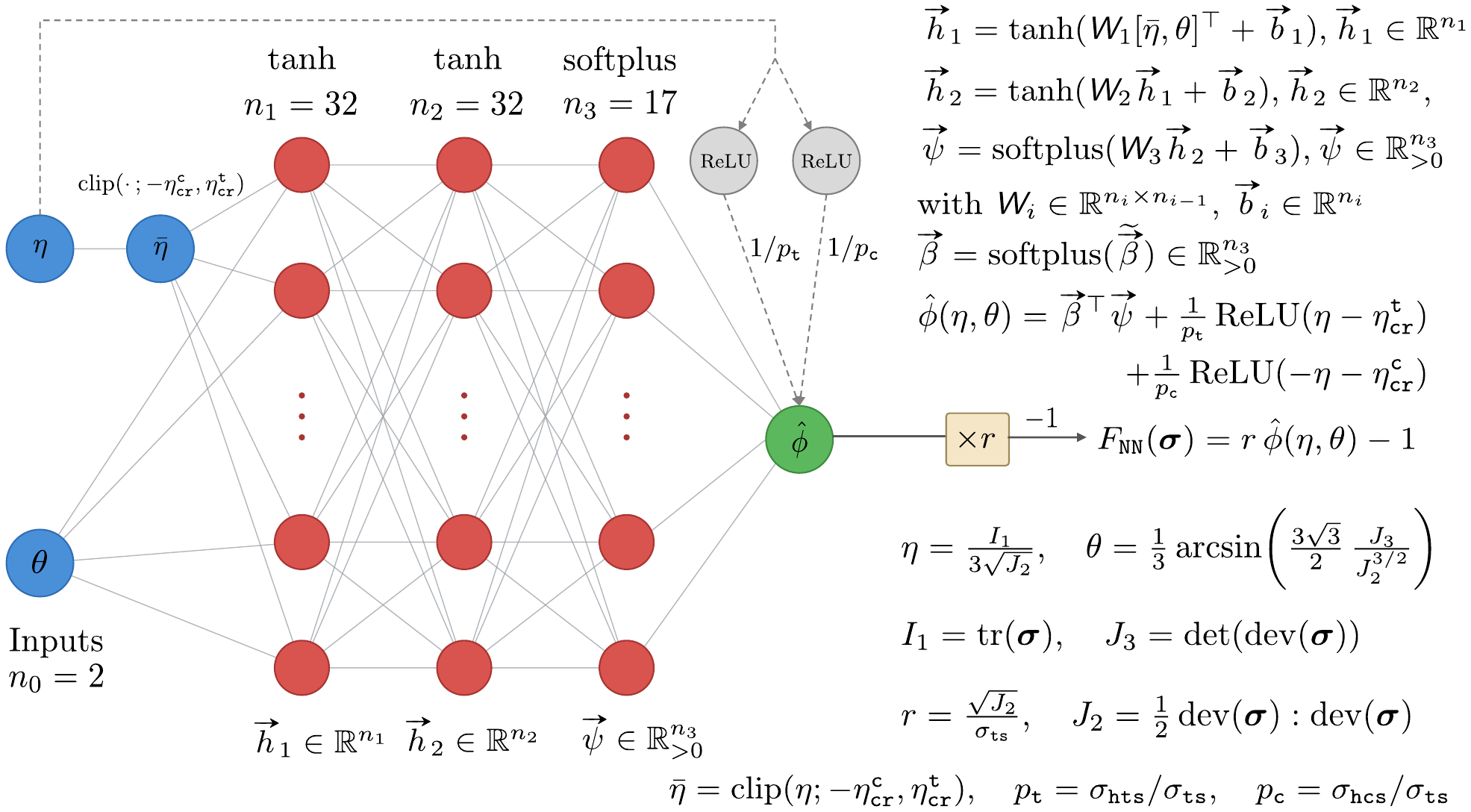}
  \caption{Neural-network strength function architecture. Inputs are the stress invariants $(\Itriax,\Lode)$---scaled triaxiality and Lode angle---defined in~\cref{eq:Invariants,eq:eta_theta}. Output is the value of the strength function. Star-convexity of the strength function with respect to the origin is ensured automatically by network construction. The scaled triaxiality is passed to a multilayer perceptron through the clip node $\etabar=\clip(\Itriax;-\etacrc,\etacrt)$, which, along with the two ReLU neurons, helps resolve finite hydrostatic strengths for hydrostatic loadings.}%
  \label{fig:nn_architecture}
\end{figure}

We first construct a neural-network architecture for the strength function. Networks of many different architectures and levels of complexity could be constructed; the one presented here was chosen after extensive numerical exploration for its ability to represent robustly all the diverse materials considered in this manuscript while avoiding over-parameterization.

\runin{Input arguments} For an isotropic material the strength function depends on the stress $\bm\sigma$ only through its  invariants $(I_1,J_2,J_3)$~\cite{KumarYavari2026starconvexity}. We use the scaled triaxiality $\Itriax$ and the Lode angle $\Lode$ of~\cref{eq:eta_theta}, together with the normalized magnitude of the deviatoric stress
\begin{equation}
r=\frac{\sqrt{J_2}}{\sts}.\label{eq:r_def}
\end{equation}
Whenever $J_2\neq0$ (non-hydrostatic loading), the triple $(r,\Itriax,\Lode)$ determines $(I_1,J_2,J_3)$ and vice versa, so any strength function of $(r,\Itriax,\Lode)$ is isotropic by construction. Along any proportional-loading ray $\bm\sigma=\lVert\bm\sigma\rVert\,\hat{\bm\sigma}$, $r$ is positively homogeneous of degree one, $r(\bm\sigma)=\lVert\bm\sigma\rVert\,r(\hat{\bm\sigma})$. For non-hydrostatic loading rays ($\hat{\bm\sigma}\neq\pm\bfI/\sqrt{3}$), $\Itriax$ and $\Lode$ are constant and independent of $\lVert\bm\sigma\rVert$. On the hydrostatic loading axis ($\hat{\bm\sigma}=\pm\bfI/\sqrt{3}$), $J_2=0$, so that $r=0$ for all $\lVert\bm\sigma\rVert$, while $\Itriax=\pm\infty$ and $\Lode$ is undefined; the treatment along this axis is elaborated later in the section.

\runin{Form of the strength function} We construct the NN strength function $\FNN$ in the form
\begin{equation}
\FNN(\bm\sigma)=  \FNN(r, \Itriax,\Lode)  = r\,\phihat(\Itriax,\Lode)-1,\qquad \phihat>0,\label{eq:FNN_form}
\end{equation}
with the loading-direction-dependent factor
\begin{equation}
\phihat(\Itriax,\Lode)={\phi}(\etabar,\Lode)+\frac{1}{\pht}\ReLU(\Itriax-\etacrt)+\frac{1}{\phc}\ReLU(-\Itriax-\etacrc),\quad {\phi}(\etabar,\Lode) = \myvect{\beta}^{\top}\myvect{\psi}(\etabar,\Lode), ~~
\etabar=\clip(\Itriax;-\etacrc,\etacrt).\label{eq:hydrofix}
\end{equation}
The first term ${\phi}(\etabar,\Lode)$ is defined through a feed-forward multilayer perceptron described below (see \textit{Network architecture}), whose positivity is enforced so that $\phi>0$; the other two terms are fixed-weight ReLU neurons where $\pht=\shts/\sts$ and $\phc=\shcs/\sts$ are the hydrostatic tensile and compressive strengths $\shts$, $\shcs$ normalized by the tensile strength $\sts$, and $\etacrt,\etacrc>0$ are cut-off values of the scaled triaxiality (see \Cref{fig:nn_architecture}). For $-\etacrc\le\Itriax\le\etacrt$ the clip is inactive ($\etabar=\Itriax$) and both ReLU terms vanish, so that $\phihat(\Itriax,\Lode)={\phi}(\etabar,\Lode)$. Closer to the hydrostatic axis ($\Itriax>\etacrt$ or $\Itriax<-\etacrc$), the clipped network input $\etabar$ is held at the cut-off and $\phihat$ transitions continuously to a function that predicts the finite hydrostatic tensile and compressive strengths $\shts$, $\shcs$ along the hydrostatic loading axis, as shown shortly (see \textit{Hydrostatic limit}).\\

Since $r\,\Itriax=I_1/(3\sts)$, multiplying~\cref{eq:hydrofix} by $r$ and substituting into \cref{eq:FNN_form} yields the following form of the strength function,
\begin{equation}
\FNN(\bm\sigma)=g_{\mathtt{NN}}(\bm\sigma)-1,\quad g_{\mathtt{NN}}(\bm\sigma)=r\,{\phi}(\etabar,\Lode)+\frac{1}{\pht}\ReLU\big(\pbar-\etacrt\,r\big)+\frac{1}{\phc}\ReLU\big(-\pbar-\etacrc\,r\big),\quad \pbar=r\,\Itriax=\frac{I_1}{3\sts},\label{eq:FNN_full}
\end{equation}
which is used in computations and remains well defined on the hydrostatic axis $\bm\sigma=\pbar\,\sts\bfI$ where $r=0$. For $-\etacrc\le\Itriax\le\etacrt$, we have $\etabar=\Itriax$ and $\FNN=r\,\phi(\Itriax,\Lode)-1$. Every term of $g_{\mathtt{NN}}$ is positively homogeneous of degree one in $\bm\sigma$ ($r$ and $\pbar$ are of degree one, $\etabar$ and $\Lode$ of degree zero), so that \emph{the NN strength function is star-convex with respect to the origin by construction}, for any values of the network parameters, by Remark~\ref{rem:homog}. %

\runin{Network architecture} The term ${\phi}(\etabar,\Lode)$ in the strength function in~\cref{eq:FNN_full} is produced by a feed-forward multilayer perceptron with layer widths $(n_0,n_1,n_2,n_3)=(2,32,32,17)$ and inputs $(\etabar,\Lode)$ (\Cref{fig:nn_architecture}):
\begin{equation}
\begin{aligned}
\myvect{h}_1 &= \tanh(\mymatrix{W}_1[\etabar,\Lode]^\top + \myvect{b}_1), & \myvect{h}_1\in\mathbb{R}^{n_1},\\
\myvect{h}_2 &= \tanh(\mymatrix{W}_2\myvect{h}_1 + \myvect{b}_2), & \myvect{h}_2\in\mathbb{R}^{n_2},\\
\myvect{\psi} &= \softplus(\mymatrix{W}_3\myvect{h}_2 + \myvect{b}_3), & \myvect{\psi}\in\mathbb{R}^{n_3}_{>0},\\
\phi(\etabar,\Lode) &= \myvect{\beta}^{\top}\myvect{\psi}, &\myvect{\beta} = \softplus(\widetilde{\myvect{\beta}})\in\mathbb{R}^{n_3}_{>0},
\end{aligned}\label{eq:NN_arch}
\end{equation}
with $\mymatrix{W}_i\in\mathbb{R}^{n_i\times n_{i-1}}$, $\myvect{b}_i\in\mathbb{R}^{n_i}$, and $\softplus(x)\equiv\ln(1+e^{x})$. The trainable parameters are $\vartheta = \{\mymatrix{W}_1,\myvect{b}_1,\allowbreak\mymatrix{W}_2,\myvect{b}_2,\allowbreak\mymatrix{W}_3,\myvect{b}_3,\allowbreak\widetilde{\myvect{\beta}}\}$. The weights $\mymatrix{W}_i$ and biases $\myvect{b}_i$ are unconstrained, while the softplus activations enforce $\myvect{\psi}>\myvect{0}$ and $\myvect{\beta}>\myvect{0}$, so that $\phi>0$. Since the ReLU terms in \cref{eq:hydrofix} are non-negative, $\phihat\ge\phi>0$; the consequences of this positivity for the  behavior of the learned surface are discussed below. The ReLU neurons and their parameters $\etacrt,\etacrc,\pht,\phc$ are directly chosen and not trained. 

\runin{Choice of architecture} The strength function in~\cref{eq:FNN_full} is the general definition of the NN strength function, whereas~\cref{eq:NN_arch} is the specific network architecture producing $\phi$ that is used in the remainder of this paper. Its hyperparameters are architectural, namely the input invariants, the number of hidden layers, the layer widths $n_1,n_2,n_3$ and the activation functions, and training-related, namely the optimizer settings, the number of epochs, and the number and multiplier $\mf$ of the fake points described below; the final-layer width $n_3$ also sets the number of output coefficients $\myvect{\beta}$ available for the PFSS calibration. The widths $n_1=n_2=32$ and $n_3=17$ were arrived at by experimentation across all materials studied here, including the requirement that the PFSS remain accurate up to $\eps/\lch\approx5$--$10$.  The triaxiality--Lode-angle basis is widely used in plasticity and failure analysis and is physically motivated. The single architecture in~\cref{eq:NN_arch} accurately captured all the synthetic strength surfaces and the experimental strength data considered in \Cref{sec:results_learning}; we therefore do not pursue a systematic hyperparameter or ablation study in this work.

\runin{Closed directions of the NN strength surface} By~\cref{eq:FNN_form,eq:FNN_full}, $g_{\mathtt{NN}}(\bm\sigma)=r\,\phihat(\Itriax,\Lode)$, so that along any proportional-loading ray $\bm\sigma=\lVert\bm\sigma\rVert\,\hat{\bm\sigma}$,
\begin{equation}
\FNN(\bm\sigma)=\lVert\bm\sigma\rVert\;g_{\mathtt{NN}}(\hat{\bm\sigma})-1,\label{eq:strict_monotone}
\end{equation}
and by Remark~\ref{rem:homog} the surface is closed along directions with $g_{\mathtt{NN}}(\hat{\bm\sigma})>0$. For non-hydrostatic rays ($\hat{\bm\sigma}\neq\pm\bfI/\sqrt{3}$), $g_{\mathtt{NN}}(\hat{\bm\sigma})=r(\hat{\bm\sigma})\,\phihat>0$, while on the hydrostatic tension and compression rays ($\hat{\bm\sigma}=\pm\bfI/\sqrt{3}$), $r=0$ and $\pbar(\hat{\bm\sigma})=\pm1/(\sqrt{3}\,\sts)$, so that by~\cref{eq:FNN_full} $g_{\mathtt{NN}}(\hat{\bm\sigma})=\pbar(\hat{\bm\sigma})/\pht=1/(\sqrt{3}\,\pht\,\sts)>0$ and $g_{\mathtt{NN}}(\hat{\bm\sigma})=-\pbar(\hat{\bm\sigma})/\phc=1/(\sqrt{3}\,\phc\,\sts)>0$, respectively. The learned surface is therefore closed along every non-hydrostatic direction, and along the hydrostatic axis it is closed with the finite strengths $\shts$, $\shcs$, being open only if the corresponding branch is switched off ($\pht$ or $\phc=\infty$).
We enforce this closedness deliberately as the default behavior: in directions not covered by training strength data, the loss function (discussed below in~\cref{eq:loss})  for training places no constraint on $\phihat$, and a network whose $\phihat$ could change sign might produce spurious open directions there, i.e.\ predict that the material never fails. With $\phihat>0$, the learned surface remains closed along all directions lacking strength data, and open directions, where physically expected, are instead enforced by choice, through the far-field points in the training loss~\cref{eq:loss} and the QP calibration~\cref{eq:qp} (which keep the surface open up to a large stress magnitude), or along the hydrostatic axis by switching off a branch. The construction also makes $\FNN$ strictly increasing along every loading ray that is not an open direction, a stronger property than star-convexity, which only requires at most a single change of sign of $\FNN$; this stronger monotonicity condition is easy to enforce and did not limit the fitting quality for any of the materials studied here.

\runin{Hydrostatic limit} Note that without the triaxiality clip and the additional ReLU neurons ($\etacrt=\etacrc=\pht=\phc=\infty$), the network would predict no strength failure under hydrostatic tensile and compressive loadings, by construction. This is because, as the hydrostatic axis is approached, $r\to0$ while $\phi$ tends to finite constants $\phi_\pm$ as $\Itriax\to\pm\infty$ (the $\tanh$ layers saturate), so that $\FNN=r\,\phi-1\to-1$. However, with the added correction neurons and clipped triaxiality, the first term in~\cref{eq:FNN_full} vanishes on the hydrostatic axis ($r=0$) since $\myvect{\beta}^{\top}\myvect{\psi}$ is bounded in~\cref{eq:NN_arch}, and~\cref{eq:FNN_full} reduces to $\FNN=\pbar/\pht-1$ for $\pbar>0$ and $\FNN=-\pbar/\phc-1$ for $\pbar<0$, with the function varying continuously across off-hydrostatic loading directions. Thus failure on the hydrostatic axis occurs at $\pbar=\pht$ in tension, and at $\pbar=-\phc$ in compression, which correspond to the hydrostatic tensile and compressive strengths $\shts$, $\shcs$, respectively.  \\

The cut-off triaxialities $\etacrt,\etacrc$ should ideally be chosen such that the interval $[-\etacrc,\etacrt]$ contains the triaxialities of all non-hydrostatic strength data, so that these data are captured by the uncorrected network. For example, experimental strength data for several materials are often restricted to the plane-stress space; every plane-stress state $\bm{\sigma}=\sigma_1\bfe_1\otimes\bfe_1+\sigma_2\bfe_2\otimes\bfe_2$ satisfies $\Itriax^2=(\sigma_1+\sigma_2)^2/\big(3(\sigma_1^2-\sigma_1\sigma_2+\sigma_2^2)\big)\le 4/3$, i.e.\ $|\Itriax|\le 2/\sqrt{3}\approx 1.155$, with equality only under equibiaxial loading. For the plane-stress data of this work we choose $\etacrt,\etacrc>2/\sqrt{3}$, such that all the results for plane-stress loading, including the training, the calibration, PFSS and finite-element results reported here, are identical with or without the hydrostatic correction neurons. Further note that hydrostatic tensile and compressive strengths can be independently prescribed in our architecture; in this work the compression branch is switched off ($\etacrc=\phc=\infty$), and the chosen $\etacrt$ and $\pht$ are listed in \ref{app:calibration}.\\

\runin{$\myvect{\beta}$-linearity and strength calibration on PFSS} As noted in Remark~\ref{rem:beta_linearity}, the closed-form solution in~\cref{eq:deltabetaeps_sol} and its generalization in~\cref{eq:lsq,eq:qp} require $F$ to be linear in its parameter vector. The network in~\cref{eq:NN_arch} is nonlinear in $\mymatrix{W}_i$, $\myvect{b}_i$ and $\widetilde{\myvect{\beta}}$, but $\FNN$ is exactly linear in the final-layer output weights $\myvect{\beta}=\softplus(\widetilde{\myvect{\beta}})$ and in the coefficients $c_\mathtt{t}\equiv1/\pht$ and $c_\mathtt{c}\equiv1/\phc$ of the two ReLU neurons. Collecting these in the parameter vector $\myvect{\beta}_{\mathtt{NN}}=[\myvect{\beta};\,c_\mathtt{t};\,c_\mathtt{c}]\in\mathbb{R}^{n_3+2}$, 
\begin{equation}
\FNN(\bm{\sigma};\myvect{\beta}_{\mathtt{NN}}) = r\,\myvect{\beta}^{\top}\myvect{\psi}(\etabar,\Lode)+c_\mathtt{t}\,\ReLU\big(\pbar-\etacrt\,r\big)+c_\mathtt{c}\,\ReLU\big(-\pbar-\etacrc\,r\big)-1,\label{eq:NN_beta_linear}
\end{equation}
which is of the form in~\cref{eq:SSurf-0} with $g$ linear in $\myvect{\beta}_{\mathtt{NN}}$. The calibration of \Cref{sec:ce,sec:generalized_calib} therefore applies unchanged, with $\myvect{\beta}$ replaced by $\myvect{\beta}_{\mathtt{NN}}$ and all other network parameters and the cut-offs held fixed at their trained and chosen values. Because the ReLU terms vanish for $-\etacrc\le\Itriax\le\etacrt$ while the first network term in~\cref{eq:NN_beta_linear} vanishes on the hydrostatic axis ($r=0$), calibration strength states of the first kind (with $-\etacrc\le\Itriax\le\etacrt$) involve only $\myvect{\beta}$, and hydrostatic calibration strength states only $c_\mathtt{t}$ and $c_\mathtt{c}$; when all calibration states are of these two kinds, the system $\mymatrix{A}\,\Dbe=\myvect{b}$ decouples into the corresponding two blocks (states with $|\Itriax|$ greater than the cut-off values but off the hydrostatic axis would couple them, and are handled by the same general solution). For example, including the hydrostatic tensile strength state $\shts\bfI$ in the calibration set contributes the row $\partial\FNN/\partial c_\mathtt{t}=\pht$ to $\mymatrix{A}$ and $-2W(\shts\bfI)/\omega_\eps$ to $\myvect{b}$ in~\cref{eq:Ab_def}, giving $c_\mathtt{t}^\eps=c_\mathtt{t}+\Delta c_\mathtt{t}^\eps=\big(1-2W(\shts\bfI)/\omega_\eps\big)/\pht$, so that the PFSS also fails at $\shts$ for tensile hydrostatic loading. If no calibration strength state lies beyond the cut-offs ($\Itriax>\etacrt$ or $\Itriax<-\etacrc$), as for all plane-stress calibrations in this work, the minimum-norm solution of~\cref{eq:lsq} leaves $c_\mathtt{t}$ and $c_\mathtt{c}$ unchanged, and a switched-off branch (such as $\phc=\infty$, $c_\mathtt{c}=0$) simply drops out. Thus the maximum number of independent stress states that can be calibrated \textit{exactly} on the PFSS is $n=n_3+2=19$, which is an order of magnitude more than the $n=2$ of the analytic surfaces. This high parameterization is what enables the accurate PFSS match at finite $\eps$ for the NN strength functions, shown later in \Cref{sec:results_learning}.

\runin{Training loss} The network is trained against $N_\texttt{data}$ strength data points $\bm{\sigma}_i$ (each known to lie on the material strength surface $\mathcal{F}$) plus, when the surface is open in some directions, $N_\texttt{fake}$ synthetic far-field points $\bm{\sigma}_{\texttt{f}j}$ placed along the open directions to enforce $\FNN\le0$ there. The loss to be minimized is
\begin{equation}
\mathcal{L}(\vartheta) = \frac{1}{N_\texttt{data}+N_\texttt{fake}}\!\left[
\sum_{i=1}^{N_\texttt{data}} \FNN(\bm{\sigma}_i;\vartheta)^2
+ \sum_{j=1}^{N_\texttt{fake}} \max(\FNN(\bm{\sigma}_{\texttt{f}j};\vartheta),0)^2
\right].\label{eq:loss}
\end{equation}
The first term forces $\FNN=0$ at the strength points; the second is a one-sided penalty that only activates when the network erroneously predicts failure ($\FNN>0$) at a far-field point along an open direction.
{Throughout this work, both the fake strength points used in learning open directions and the far-field strength probe points used in calibrating these open directions in the PFSS (using~\cref{eq:qp}) are placed along the open directions at the stress magnitude $\lVert\bm{\sigma}_{\texttt{f}j}\rVert=\mf\max_i\lVert\bm{\sigma}_i\rVert$, i.e.\ a multiple $\mf$ of the largest stress magnitude in the training data; $\mf=10$ in training, while the calibration multiplier is chosen per material (\ref{app:calibration}).}

\runin{Optimization} We train the network using the Adam optimizer~\cite{KingmaBa2014Adam} with learning rate $\alpha = 2\times 10^{-3}$ and Adam hyperparameters $\beta^{\texttt{Adam}}_1 = 0.9$, $\beta^{\texttt{Adam}}_2 = 0.999$, $\eps_\texttt{Adam} = 10^{-8}$. Gradients $\nabla_\vartheta\mathcal{L}$ are computed by reverse-mode automatic differentiation. The fake-point contribution to the gradient activates only when $\FNN(\bm{\sigma}_{\texttt{f}j})>0$. We use the PyTorch~\cite{PyTorch2019} implementation as the primary  implementation; an independent NumPy implementation was also run as a cross-check, and for each case the better of the two fits (by final loss and smoothness of the learned surface) is retained: the NumPy fit for Mohr--Coulomb and the PyTorch fit for all other cases. Weights $\mymatrix{W}_i$ and biases $\myvect{b}_i$ are initialized from the PyTorch default uniform distribution $\mathcal{U}(-1/\sqrt{n_{i-1}},+1/\sqrt{n_{i-1}})$; the NumPy implementation draws $\mymatrix{W}_i$ from the same distribution but initializes $\myvect{b}_i=\myvect{0}$. The output coefficients are initialized as $\widetilde{\myvect{\beta}}=\myvect{0}$ so that initial $\myvect{\beta}=\softplus(\myvect{0})=\ln 2$.

\runin{Stopping criterion} Adam optimization in this setting exhibits oscillatory but decreasing loss at large epochs (visible in the loss panels of \Cref{fig:syn_hb,fig:syn_mc}). We track the lowest-loss parameter snapshot $\vartheta_\texttt{best}$ encountered during training and use it for all post-training PFSS evaluations. For the synthetic-data cases (\Cref{sec:syn_results}) the maximum-epoch budget is $5\times 10^4$, well beyond loss convergence. For experimental data (\Cref{sec:exp_results}), where the data are noisy and unevenly spaced, we stop training earlier (typically $8\text{--}9\times 10^3$ epochs) at the point where a good fit has been achieved but before the loss further descends into a regime that produces wavy, overfit surfaces. The stopping epoch is a customizable judgment call, made per material and reported in \ref{app:calibration}. A more quantitative alternative is early stopping based on a held-out validation set, in which part of the data is withheld from training and training is stopped once the loss on the withheld data starts to increase; since the experimental data sets considered here are small, we did not withhold data for validation and instead selected the stopping epoch from visual inspection of the learned surfaces.

\subsection{Analytic strength surfaces}\label{sec:analytic_F}

The four analytic strength surfaces below appear in two distinct roles in this work (\Cref{sec:syn_results}): (i)~to show the ability of the NN strength function to capture the most popular analytic strength surfaces; and (ii)~as low-parameter (but current state-of-the-art) benchmarks against which the PFSS of the NN strength function is compared at finite $\eps$. All four surfaces are linear in their $n=2$ parameter vector $\myvect{\beta}=[\beta_1,\beta_2]$ by construction. %

\subsubsection{Mohr--Coulomb}\label{sec:mc}

The Mohr--Coulomb (M--C) surface~\cite{labuz2014mohr} is the linear strength function
\begin{equation}
\mathcal{F}_\mathtt{MC} \equiv \Fmc(\bm{\sigma};\myvect{\beta}) = \beta_1\sigmax + \beta_2\sigmin - 1 = 0,\qquad
\beta_1 = \frac{1}{\sts},\;\beta_2 = -\frac{1}{\scs}.\label{eq:F_MC}
\end{equation}
$\Fmc$ depends only on the maximum and minimum principal stresses; the intermediate principal stress plays no role.

\subsubsection{Mogi--Coulomb}\label{sec:mgc}

The Mogi--Coulomb (Mg--C) surface~\cite{mogi1971fracture,al2005relation} is the linear strength function
\begin{equation}
\mathcal{F}_\mathtt{MgC} \equiv \Fmgc(\bm{\sigma};\myvect{\beta}) = \beta_1\,\smTwo + \beta_2\,\toct - 1 = 0,\qquad
\beta_1 = \frac{1}{\scs} - \frac{1}{\sts},\;\beta_2 = \frac{3\sqrt{2}}{4}\!\left(\frac{1}{\sts}+\frac{1}{\scs}\right),\label{eq:F_MgC}
\end{equation}
with $\smTwo = -\sigmTwo$ and $\toct = \sqrt{\tfrac{2}{3}J_2}$. The Mg--C surface uses all three principal stresses but reduces to the M--C surface in loadings where there is no distinct intermediate principal stress.

\subsubsection{3D Hoek--Brown}\label{sec:hb}

The 3D Hoek--Brown (3D H--B) surface in the generalized Zhang--Zhu form~\cite{zhang2007three,zhang2008generalized,Chockalingam2026EFM} is the  strength function
\begin{equation}
\mathcal{F}_\mathtt{HB} \equiv \Fhb(\bm{\sigma};\myvect{\beta}) =
\beta_1\!\left(\frac{3}{\sqrt{2}}\,\toct\right)^{\!1/a} +
\beta_2\!\left(\frac{3}{2\sqrt{2}}\,\toct - \smTwo\right) - 1 = 0,\label{eq:F_HB}
\end{equation}
with
\begin{equation}
\beta_1 = \frac{1}{\scs^{1/a}},\qquad
\beta_2 = \frac{1}{\sts}\!\left(1-\!\left(\frac{\sts}{\scs}\right)^{\!1/a}\right),\label{eq:beta_HB}
\end{equation}
where $a\in(0,1]$ is a Hoek--Brown rock-mass parameter; for intact rock $a=1/2$, which is the value used here. Among the strength functions considered in this paper, the 3D H--B {strength function} is the only one {that is non-homogeneous in stress}.

\subsubsection{Drucker--Prager}\label{sec:dp}

The Drucker--Prager (D--P) surface~\cite{drucker1952soil} is the linear strength function
\begin{equation}
\mathcal{F}_\mathtt{DP} \equiv \Fdp(\bm{\sigma};\myvect{\beta}) = \beta_1\,I_1 + \beta_2\,\sqrt{J_2} - 1 = 0,\qquad
\beta_1 = \frac{1}{2}\!\left(\frac{1}{\sts}-\frac{1}{\scs}\right),\;\beta_2 = \frac{\sqrt{3}}{2}\!\left(\frac{1}{\sts}+\frac{1}{\scs}\right).\label{eq:F_DP}
\end{equation}

The D--P surface is open along the hydrostatic compression axis whenever $\scs>\sts$, and in plane stress along certain compression--compression directions  whenever $\scs > 3\,\sts$~\cite{chockalingam2025construction}, which is the case for many materials (e.g.\ ceramics including titania). Where the D--P surface is open, training-data fitting and PFSS calibration must respect this openness; this is enforced through the far-field fake points in the training loss and the QP calibration of \Cref{sec:generalized_calib}.\\

{This completes the description of the strength functions. In the next section, we demonstrate the ability of the NN strength function to learn both synthetic strength data generated from these analytic strength surfaces and experimental strength data, and to be incorporated accurately into the phase-field theory across regularization lengths, in comparison with the exact analytic strength functions.}

\section{Results: learning strength from data and PFSS}\label{sec:results_learning}

We now demonstrate the ability of the NN strength function constructed in~\Cref{sec:strength_functions} both to capture a wide range of synthetic and experimental strength data across  various materials and to accurately incorporate the learned surfaces into the phase-field theory across regularization lengths. We first consider, in \Cref{sec:syn_results}, synthetic data generated from the  four analytic strength surfaces discussed in \Cref{sec:analytic_F}: 3D Hoek--Brown, Mohr--Coulomb, Drucker--Prager, and Mogi--Coulomb. We then consider experimental strength data in \Cref{sec:exp_results} for graphite, natural rubber, and styrene--butadiene rubber. Throughout this section, all results are presented in the plane-stress principal-stress space ($\sigma_3=0$), which is the most common testing condition; the framework itself is fully three-dimensional and some representative results for full 3D loading are presented in \Cref{sec:exp_3d}. For all PFSS plots, the scalar $\de$ is set to unity except for graphite, where $\de$ has been calibrated separately at each $\eps/\lch$ via the single-edge notched tension (SENT) test described in \Cref{sec:fea_sent}. The per-material training and calibration details, such as the numbers of calibration strength states, the openness directions and the multiplier $\mf$, are reported in \ref{app:calibration}; the calibration locations are given in the respective subsections.

\runin{Calibration strengths for PFSS construction} For synthetic-data cases,  the training data lie exactly on the analytic surface, and  the strength locations $\bm{\sigma}_{\texttt{s}i}$ at which the PFSS calibration {in}~\cref{eq:lsq,eq:qp} is enforced are chosen to coincide with the training data. In experimental-data cases, the training data are noisy and unevenly spaced; thus we instead choose $\bm{\sigma}_{\texttt{s}i}$ to be stress states located on the \emph{learned} surface $\FNN=0$ along equiangularly distributed loading rays. In this way, the PFSS calibration is not affected by the noise of individual data points.

\subsection{Synthetic data from analytic functions}\label{sec:syn_results}

For the analytic strength surfaces, we generate synthetic strength data through periodic sampling in the stress space for the  NN strength function to learn. These data are regular and noise-free; the focus here is to demonstrate that a single NN strength function captures all the different analytic strength surfaces. Real, noisy experimental data are considered in \Cref{sec:exp_results}. For each analytic strength surface (3D H--B, M--C, D--P and Mg--C), we sample $N_\texttt{data} = 32$ strength states in principal-stress space ($\sigma_1$--$\sigma_2$) that are equiangularly distributed on the strength surface with respect to the origin. Due to the $\sigma_1\leftrightarrow\sigma_2$ symmetry, up to $17$ of these are distinct. In open directions (only D--P  among these four surfaces has open directions), training data are skipped and replaced by far-field fake training points {$\bm{\sigma}_{\texttt{f}j}$} with $\mf=10$ to enforce openness via the one-sided loss in~\cref{eq:loss}.  The 3D H--B, M--C, and Mg--C surfaces are generated using Indiana Limestone parameters from~\cite{Chockalingam2026EFM}; the D--P surface is generated using Titania parameters from~\cite{KBFLP20,ely1972strength,iuga2007ab}. These material parameters are listed in \ref{app:materials}.

\subsubsection{3D Hoek--Brown}\label{sec:syn_hb}

\begin{figure}[h]
  \centering
  \includegraphics[width=\textwidth]{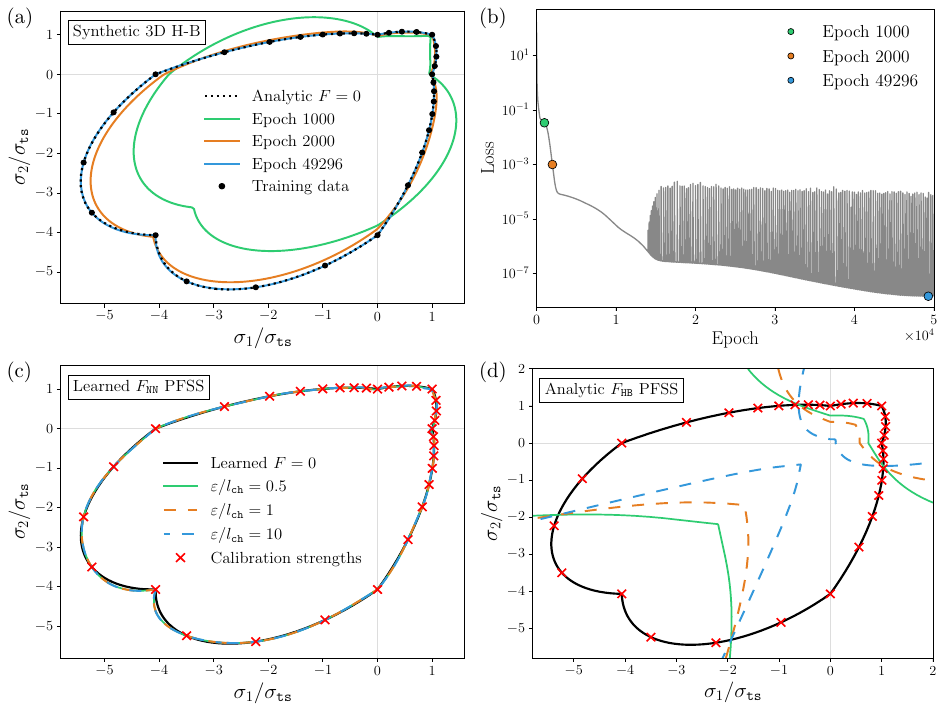}
  \caption{Machine learning and phase-field incorporation of the 3D Hoek--Brown strength surface (for Indiana Limestone parameters, see~\ref{app:materials}), in the plane-stress principal-stress space. (a) The neural-network strength surface $\FNN=0$ progressively learning the discrete synthetic strength data (large black dots) on the full analytic 3D H--B strength surface $\Fhb=0$ (dotted lines, \cref{eq:F_HB}), with increasing training epochs. The learned strength surface $\FNN=0$ is shown at different training epochs in solid colored lines; the blue line is the final learned surface at the end of training (lowest-loss epoch). (b) Training loss vs.\ epoch; markers indicate the chosen epochs shown in (a). (c)--(d) The phase-field strength surface $\mathcal{F}^\eps$ (\cref{eq:PFSS_general}) obtained upon incorporation of (c) the learned surface $\FNN=0$ and (d) the analytic surface $\Fhb=0$, for varying regularization lengths. The surfaces being incorporated are shown in black. The calibration strength states are marked using red crosses.}
  \label{fig:syn_hb}
\end{figure}
  
 The results for machine learning of the NN strength surface $\FNN=0$ using the synthetic plane-stress data for the 3D H--B surface are shown in \Cref{fig:syn_hb}(a). The dotted lines represent the true analytic strength surface $\Fhb=0$ defined in \cref{eq:F_HB} and the large black dots represent the discrete, equiangularly spaced training data sampled on this surface. The trained surfaces are shown using colored lines for increasing epochs---the progressive learning of the synthetic strength data can be seen, with the  final best-epoch learned surface visually indistinguishable from the analytic surface---even between training points. Thus, the NN strength function is able to capture the 3D H--B strength surface extremely well. Note that the training loss {in principle only forces the NN strength surface to capture the training data points; however, the network construction here also recovers the underlying analytic strength surface.} \Cref{fig:syn_hb}(b) plots the corresponding evolution of the training loss, which exhibits oscillatory but generally decreasing behavior characteristic of the Adam optimizer.

\runin{Excellent PFSS for $\FNN$} \Cref{fig:syn_hb}(c) plots the PFSS obtained upon the phase-field incorporation of the learned strength surface (shown as a solid black line) at various values of the regularization length $\eps$, using $F=\FNN$ in \cref{eq:PFSS_general}, with the calibration strength locations marked using red crosses. Note that a maximum of 17 independent strength locations can be captured exactly on the PFSS in plane-stress loading for our chosen form of the $\FNN$ (since $n_3=17$ for $\myvect{\beta}$ in $\FNN$), and more can be captured approximately in a least-squares sense as described in \Cref{sec:generalized_calib}. While the calibration strength states have been chosen here to coincide with the training data, they can be independently chosen in general. It is seen in \Cref{fig:syn_hb}(c) that across regularization lengths, even up to ten times the characteristic length $\lch$, the PFSS closely matches the learned strength surface (which was already shown to closely capture the analytic strength surface). While in principle the PFSS only needs to match the material strength surface closely at the calibration strength states and can deviate elsewhere, here even the match away from calibration states remains accurate, even at the large regularization lengths---likely because the calibration states are closely spaced, leaving little room for deviation between them. This is enabled by the high parameter dimensionality of the NN strength function, which allows more stress states on the strength surface to be calibrated during phase-field incorporation. Further, {unlike in earlier works~\cite{KRLP22,Chockalingam2026EFM}, no compression correction is required, which also avoids the spurious discontinuities that such corrections introduce in the PFSS.}  That the PFSS of the NN strength function accurately captures the learned surface across regularization lengths is a significant improvement over the current state of the art, which incorporates the exact analytic 3D H--B form into the phase-field theory; these results are shown next for comparison.

\begin{figure}[h]
  \centering
  \includegraphics[width=\textwidth]{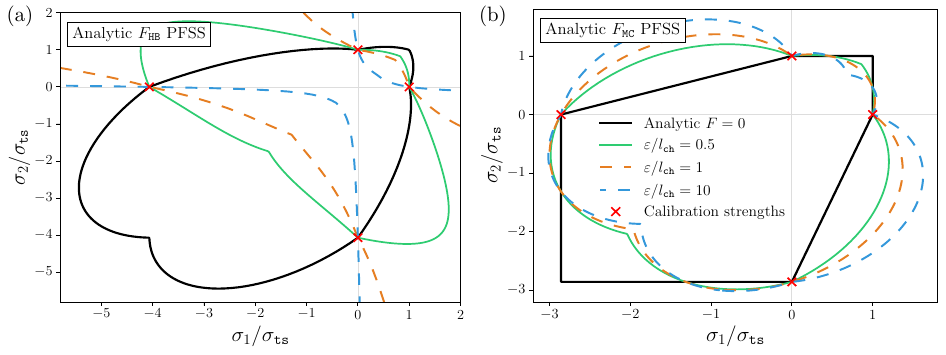}
  \caption{Phase-field incorporation of the \emph{exact analytic} (a)~3D Hoek--Brown $\Fhb=0$ (\cref{eq:F_HB}) and (b)~Mohr--Coulomb $\Fmc=0$ (\cref{eq:F_MC}) strength surfaces (for Indiana Limestone parameters, see~\ref{app:materials}), shown in the plane-stress principal-stress space. The analytic surfaces being incorporated are shown using a solid black line while the phase-field strength surface $\mathcal{F}^\eps$ (\cref{eq:PFSS_general}) is shown  for varying regularization lengths using colored lines. The calibration strength states are marked using red crosses which correspond here to the  two independent strength locations $\sts$ and $\scs$ (four total states visualized due to the $\sigma_1\leftrightarrow\sigma_2$ symmetry). }
  \label{fig:syn_hb_2calib}
\end{figure}
 
\runin{Poor PFSS for analytic $\Fhb$} The strength surfaces obtained by the phase-field incorporation of the exact analytic 3D H--B form are plotted in \Cref{fig:syn_hb}(d) and \Cref{fig:syn_hb_2calib}(a), by using $F=\Fhb$ in \cref{eq:PFSS_general}. In \Cref{fig:syn_hb}(d), the same 32 calibration strength states (17 distinct) used  in \Cref{fig:syn_hb}(c) for the NN strength function are used, whereas in \Cref{fig:syn_hb_2calib}(a) four calibration strength states (2 independent) are used instead. The latter calibration is a better choice that reflects the lower parameter dimensionality of the analytic 3D H--B form wherein a maximum of 2 independent strength states can be exactly captured in the PFSS ($n=2$ for $\myvect{\beta}$ in $\Fhb$). However, it is seen that for either choice of calibration strengths, the PFSS match with the material strength surface is poor even at the lower regularization length of $\eps=0.5\,\lch$ and gets worse with increasing $\eps$. For the choice of 2 independent calibration strengths, the PFSS, as expected, remains exact at the calibration strengths across the regularization lengths, but away from them the accuracy is poor and gets worse with increasing $\eps$. This is a central result---\textit{even when the exact strength surface that describes the material is known, incorporation into the phase-field theory results in poor reproduction of the strength surface when the parameter dimensionality of the strength function is low (as is typically the case)}, unless much smaller regularization lengths are used, at which the representation becomes accurate~\cite{Chockalingam2026EFM}. This represents a significant bottleneck for large-scale numerical simulations as discussed earlier. \textit{In comparison, due to its high parameter dimensionality, the NN strength function is not only able to learn the strength surface but also incorporate it faithfully into the phase-field theory across regularization lengths.} \\
 
 Through the rest of this subsection, the ability of the NN strength function both to learn the analytic strength surface and to incorporate it accurately into the phase-field theory is demonstrated for the remaining analytic strength surfaces. 

\subsubsection{Mohr--Coulomb}\label{sec:syn_mc}

\begin{figure}[h]
  \centering
  \includegraphics[width=\textwidth]{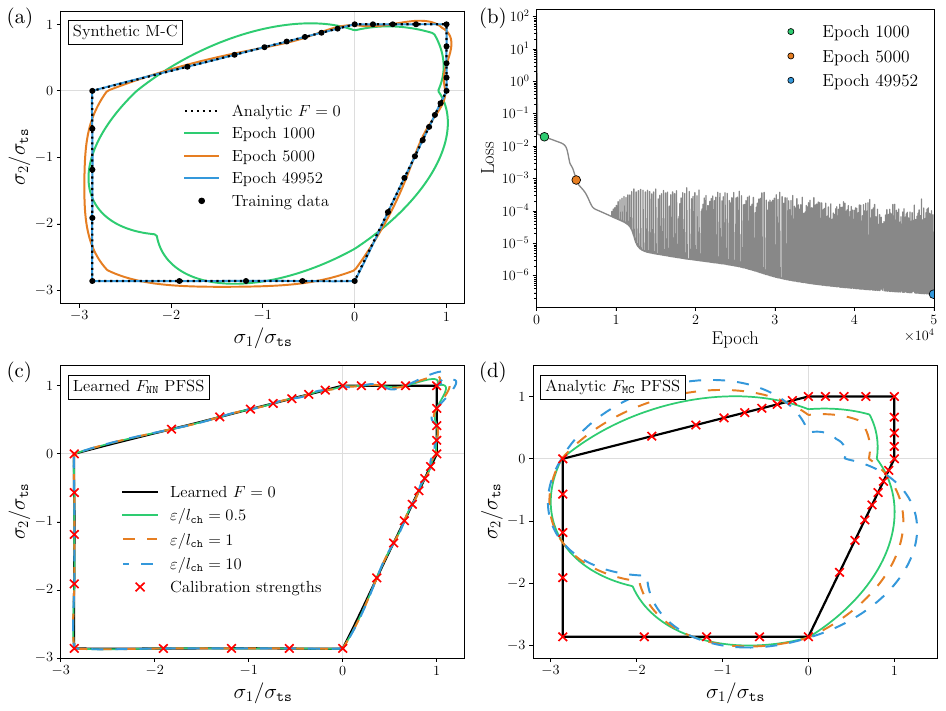}
  \caption{Machine learning and phase-field incorporation of the Mohr--Coulomb strength surface (for Indiana Limestone parameters, see~\ref{app:materials}), in the plane-stress principal-stress space. (a) The neural-network strength surface $\FNN=0$ progressively learning the discrete synthetic strength data (large black dots) on the analytic M--C strength surface $\Fmc=0$ (dotted line, \cref{eq:F_MC}), with increasing training epochs. The learned strength surface $\FNN=0$ is shown at different training epochs in solid colored lines; the blue line is the final learned surface at the end of training (lowest-loss epoch). (b) Training loss vs.\ epoch; markers indicate the chosen epochs shown in (a). (c)--(d) The phase-field strength surface $\mathcal{F}^\eps$ (\cref{eq:PFSS_general}) obtained upon incorporation of (c) the learned surface $\FNN=0$ and (d) the analytic surface $\Fmc=0$, for varying regularization lengths. The surfaces being incorporated are shown in black. The calibration strength states are marked using red crosses.}
  \label{fig:syn_mc}
\end{figure}

Following the same procedure as for the 3D H--B surface (\Cref{sec:syn_hb}), the results for machine learning of the NN strength surface $\FNN=0$ using the synthetic plane-stress data for the M--C surface are shown in \Cref{fig:syn_mc}(a)--(b). The NN strength surface progressively learns the synthetic data, and the final learned surface is visually indistinguishable from the analytic surface $\Fmc=0$ (\Cref{fig:syn_mc}(a)). Thus, the NN strength function captures the M--C strength surface extremely well, \textit{including its sharp corners}. The PFSS
obtained by incorporating the final learned strength surface into the phase-field theory is shown in \Cref{fig:syn_mc}(c). Once again, the PFSS match with the learned surface is excellent across regularization lengths. There is some minor mismatch in the tension--tension quadrant between the calibration states (at which the PFSS is exact), and this mismatch decreases with decreasing regularization length. In contrast, the PFSS obtained by incorporating the exact analytic strength function $\Fmc$ is plotted in \Cref{fig:syn_hb_2calib}(b) and \Cref{fig:syn_mc}(d), once again for 2 independent and 17 distinct calibration strength states, respectively. As before, for either choice of calibration strengths, the PFSS match with the material strength surface is poor even at the lower regularization length of $\eps=0.5\,\lch$ and gets worse with increasing $\eps$. Thus, as for the 3D H--B surface, the NN strength function both learns the M--C surface and incorporates it accurately into the phase-field theory across regularization lengths, whereas the PFSS of the exact analytic M--C surface is poor.

\subsubsection{Drucker--Prager}\label{sec:syn_dp}

\begin{figure}[h]
  \centering
  \includegraphics[width=\textwidth]{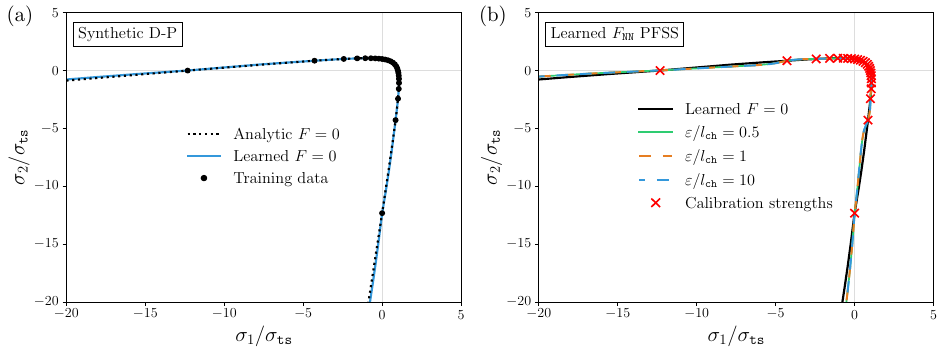}
  \caption{Machine learning and phase-field incorporation of the Drucker--Prager strength surface (for Titania parameters, see~\ref{app:materials}), in the plane-stress principal-stress space. The D--P surface is open in part of the compression--compression quadrant ($\scs/\sts\approx 12.3 > 3$). (a) The final learned neural-network strength surface $\FNN=0$ (solid line) and the analytic D--P strength surface $\Fdp=0$ (dotted line, \cref{eq:F_DP}), with the synthetic strength data (black dots); along the open directions, the training data are replaced by fake points (not shown) at $\lVert\bm{\sigma}_{\texttt{f}j}\rVert=10\max_i\lVert\bm{\sigma}_i\rVert$. (b) The phase-field strength surface $\mathcal{F}^\eps$ (\cref{eq:PFSS_general}) obtained upon incorporation of the learned surface $\FNN=0$ (black), for varying regularization lengths, calibrated using the QP in~\cref{eq:qp} with $\mf=2.5$. The calibration strength states are marked using red crosses.}
  \label{fig:syn_dp}
\end{figure}

The results for the D--P strength surface, the first case requiring openness enforcement, are shown in \Cref{fig:syn_dp}. For the Titania parameters used here, the analytic D--P surface is open along part of the compression--compression quadrant; in those directions we omit the training data (seven of the $32$ equiangular directions from the origin, $191.25^\circ$--$258.75^\circ$, leaving $N_\texttt{data}=25$) and instead place fake far-field points (with distance multiplier $\mf=10$) that enforce the openness through the one-sided penalty term in the training loss in~\cref{eq:loss}. The learned surface at the end of training is shown in \Cref{fig:syn_dp}(a) and once again closely captures the analytic surface, including its openness in the compression--compression quadrant (over the plotted range). However, as discussed in \Cref{sec:nn_F}, the NN strength surface is closed by construction along all non-hydrostatic loading directions; it therefore technically closes along the ``open'' directions as well, but at very large stress magnitudes, so that it can be considered open for practical purposes.\\

 The PFSS obtained by phase-field incorporation of the learned surface is shown in \Cref{fig:syn_dp}(b). For the PFSS, the $25$ training strength locations are used as calibration strength states, and far-field probes (with distance multiplier $\mf=2.5$, which was found to give the best results) are placed along each of the seven open directions to enforce openness through the QP solution in~\cref{eq:qp}. The PFSS closely matches the learned surface across the regularization lengths and remains open in the required directions in the compression--compression quadrant over the plotted range. Thus, the NN strength function is also able to learn the D--P surface and incorporate it accurately into the phase-field theory.

\subsubsection{Mogi--Coulomb}\label{sec:syn_mgc}

\begin{figure}[h]
  \centering
  \includegraphics[width=\textwidth]{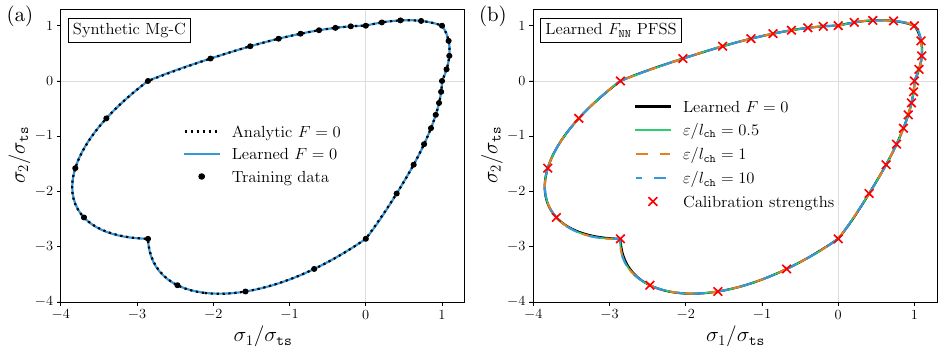}
  \caption{Machine learning and phase-field incorporation of the Mogi--Coulomb strength surface (for Indiana Limestone parameters, see~\ref{app:materials}), in the plane-stress principal-stress space. (a) The final learned neural-network strength surface $\FNN=0$ (solid line) and the analytic Mg--C strength surface $\Fmgc=0$ (dotted line, \cref{eq:F_MgC}), with the synthetic strength data (black dots). (b) The phase-field strength surface $\mathcal{F}^\eps$ (\cref{eq:PFSS_general}) obtained upon incorporation of the learned surface $\FNN=0$ (black), for varying regularization lengths. The calibration strength states are marked using red crosses.}
  \label{fig:syn_mgc}
\end{figure}

The results for the Mg--C strength surface are shown in \Cref{fig:syn_mgc}. The learned surface is again visually indistinguishable from the analytic surface $\Fmgc=0$ (\Cref{fig:syn_mgc}(a)), and its PFSS closely matches it up to $\eps=10\,\lch$ (\Cref{fig:syn_mgc}(b)). Together, \Cref{sec:syn_hb,sec:syn_mc,sec:syn_dp,sec:syn_mgc} show that a single neural-network architecture captures closed and open, smooth and cornered analytic strength surfaces, and that, owing to its high parameter dimensionality, its PFSS remains highly accurate across regularization lengths where the PFSS of even the exact analytic surfaces is poor.

\subsection{Learning from experimental data}\label{sec:exp_results}

Having demonstrated the ability of the NN strength function to learn a wide variety of analytic strength surfaces, we now demonstrate its ability to learn directly from experimental strength data. We consider plane-stress experimental strength data from two distinct material classes: nuclear-grade graphite, a linear-elastic brittle material, and natural rubber (NR) and styrene--butadiene rubber (SBR), two hyperelastic elastomers. Together, these data exhibit the features of practical experimental strength data that the NN strength function must accommodate: noisy measurements, missing quadrants, and non-convexity. For each material, the network architecture and training pipeline of \Cref{sec:nn_F} are used unchanged; only the openness-enforcement directions, the training-stopping epoch and the calibration states differ per material.

\subsubsection{Graphite}\label{sec:exp_graphite}

\begin{figure}[h]
  \centering
  \includegraphics[width=\textwidth]{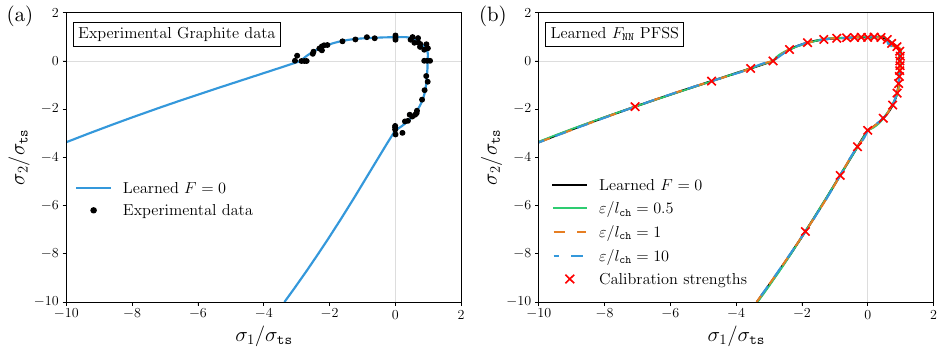}
  \caption{Machine learning and phase-field incorporation of the experimental strength data of nuclear-grade graphite~\cite{sato1987fracture}, in the plane-stress principal-stress space. (a) The learned neural-network strength surface $\FNN=0$ (solid line) and the experimental strength data (black dots). The data lie in the tension--tension and tension--compression quadrants; openness is enforced along the equibiaxial-compression direction $225^\circ$ in the compression--compression quadrant. (b) The phase-field strength surface $\mathcal{F}^\eps$ (\cref{eq:PFSS_general}) obtained upon incorporation of the learned surface $\FNN=0$ (black), for varying regularization lengths, with $\de$ calibrated at each $\eps/\lch$ using the SENT test of \Cref{sec:fea_sent}. The calibration strength states (red crosses) are placed on the learned surface rather than on the experimental data. Material parameters are listed in \ref{app:materials}.}
  \label{fig:exp_graphite}
\end{figure}

The results for graphite are shown in \Cref{fig:exp_graphite}: the strength surface learned by the NN strength function from the experimental strength data is shown in \Cref{fig:exp_graphite}(a), and its PFSS in \Cref{fig:exp_graphite}(b). The strength data~\cite{sato1987fracture,KBFLP20} (\Cref{fig:exp_graphite}(a)) lie in the tension--tension and tension--compression quadrants, but no failure data are available in the compression--compression quadrant. We are free to enforce openness in that quadrant; we enforce it along the single equibiaxial-compression direction $225^\circ$, through one fake point in training ($\mf=10$) and one far-field probe in the QP calibration. The learned surface then remains open over a range of directions about this axis within the plotted range, and transitions smoothly to the data-supported regions (\Cref{fig:exp_graphite}(a)). This flexibility to enforce openness in directions outside the data envelope is an advantage of the NN strength function that is not available with analytic forms. The learned strength surface fits the noisy experimental data smoothly (\Cref{fig:exp_graphite}(a)). Training is stopped at $8\times 10^3$ epochs: fewer epochs do not fit the data accurately, while further training produces wavy surfaces that overfit the noisy data.\\

The PFSS obtained by phase-field incorporation of the learned surface is shown in \Cref{fig:exp_graphite}(b) across regularization lengths. Since the learned graphite surface is used as the representative case in the finite-element simulations of \Cref{sec:results_fea}, we use here the values of $\de$ calibrated from the SENT test  instead of the value $\de=1$ in \cref{eq:PFSS_general} used for all other PFSS results in this section. As explained in \Cref{sec:ce}, $\de$ is calibrated such that the failure load predicted for a large pre-existing crack agrees with the Griffith criticality condition $G=G_c$ in~\cref{eq:Griffith}; the SENT test is described in \Cref{sec:fea_sent}. The calibrated values are listed in \Cref{tab:de_graphite}. For the calibration, the noisy and unevenly spaced experimental data are poor choices of calibration strength states; instead, we place $\bm{\sigma}_{\texttt{s}i}$ on the learned surface $\FNN=0$ along the $25$ of $32$ equiangular directions ($11.25^\circ$ spacing) lying outside the sector $(180^\circ,270^\circ)$, augmented by eight states along $180^\circ+t$ and $270^\circ-t$, $t\in\{5^\circ,10^\circ,15^\circ,20^\circ\}$, for $33$ calibration strength states in total; the QP in~\cref{eq:qp} uses a single far-field probe along $225^\circ$ with distance multiplier $\mf=5$. The PFSS of the learned surface closely matches it across regularization lengths, up to $\eps=10\,\lch$ (\Cref{fig:exp_graphite}(b)). The same network parameters and calibrated $\de$ values are used for all graphite finite-element simulations in \Cref{sec:results_fea}.\\

Thus, the NN strength function learns directly from the noisy experimental data of graphite while enforcing the prescribed openness, and the learned strength surface is accurately incorporated into the phase-field theory across regularization lengths.

\subsubsection{Natural rubber (NR)}\label{sec:exp_nr}

\begin{figure}[h]
  \centering
  \includegraphics[width=\textwidth]{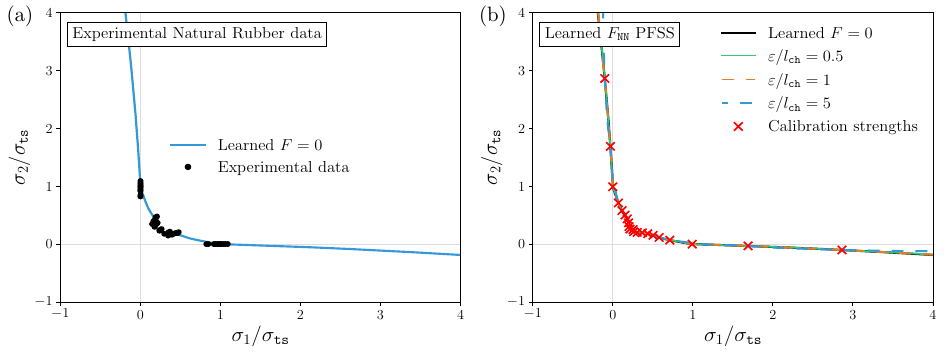}
  \caption{Machine learning and phase-field incorporation of the experimental strength data of natural rubber~\cite{hamdi2006rubber}, in the plane-stress principal-stress space. (a) The learned neural-network strength surface $\FNN=0$ (solid line) and the experimental strength data (black dots), all in the tension--tension quadrant; openness is enforced over $95^\circ$--$355^\circ$. (b) The phase-field strength surface $\mathcal{F}^\eps$ (\cref{eq:PFSS_general}) obtained upon incorporation of the learned surface $\FNN=0$ (black) with the Yeoh strain energy (\cref{eq:Yeoh}), for varying regularization lengths. The calibration strength states are marked using red crosses. Material parameters are listed in \ref{app:materials}.}
  \label{fig:exp_nr}
\end{figure}

The results for natural rubber are shown in \Cref{fig:exp_nr}: the strength surface learned by the NN strength function from the experimental strength data is shown in \Cref{fig:exp_nr}(a), and its PFSS in \Cref{fig:exp_nr}(b). The strength data~\cite{hamdi2006rubber} cover only the tension--tension quadrant; elastomers do not typically exhibit brittle compressive failure, so no failure data are available elsewhere. Moreover, the data are noisy and non-convex (see Remark~\ref{rem:notconvex}): the strengths near equibiaxial tension lie well inside (on the same side as the origin) the straight line joining the two uniaxial tensile strengths on the two axes (\Cref{fig:exp_nr}(a)). Such data cannot be captured well by the commonly used analytic strength surfaces. %
 Specialized failure criteria, typically formulated in terms of the principal stretches rather than the stresses, have therefore been developed for elastomers~\cite{hamdi2006rubber,rosendahl2019equivalent}.\\
 
 In contrast, the NN strength function, whose only built-in constraint is star-convexity with respect to the origin, can capture the non-convex experimental data well. We use the same architecture and training procedure as for the analytic surfaces in \Cref{sec:syn_results} and for graphite, and enforce openness over $95^\circ$--$355^\circ$ (16 equiangularly spaced fake points with $\mf=10$), %
 which keeps the learned surface smooth where the data end. Training is stopped at $9\times 10^3$ epochs. The learned surface closely fits the noisy, non-convex experimental data while remaining open in the enforced directions over the plotted range (\Cref{fig:exp_nr}(a)).\\

{The PFSS is obtained using~{\cref{eq:PFSS_general}} with} the hyperelastic Yeoh strain energy in~\cref{eq:Yeoh} (parameters in \ref{app:materials}). As in the graphite case, the calibration strength states are placed on the learned surface $\FNN=0$, here along the $17$ of $64$ equiangular directions ($5.625^\circ$ spacing) lying outside the open sector, plus four states along $91^\circ$, $92^\circ$, $358^\circ$ and $359^\circ$ near the edges of the open sector, $21$ in total; the QP in~\cref{eq:qp} uses 16 far-field probes equally spaced from $95^\circ$ to $355^\circ$ (both included) with $\mf=2$. The PFSS closely matches the learned surface up to $\eps=5\,\lch$ and remains open in the enforced directions over the plotted range (\Cref{fig:exp_nr}(b)); at larger $\eps/\lch$, the PFSS performs poorly, and the elastomer results are therefore reported up to $\eps=5\,\lch$. Thus, the same NN strength function learns noisy, non-convex experimental strength data with prescribed openness, and the learned surface is accurately incorporated into the phase-field theory even with a hyperelastic strain energy.

\subsubsection{Styrene--butadiene rubber (SBR)}\label{sec:exp_sbr}

\begin{figure}[!htbp]
  \centering
  \includegraphics[width=\textwidth]{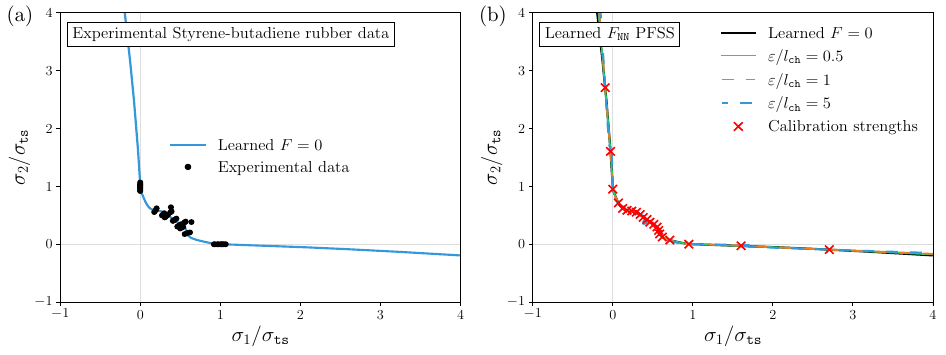}
  \caption{Machine learning and phase-field incorporation of the experimental strength data of styrene--butadiene rubber~\cite{hamdi2006rubber}, in the plane-stress principal-stress space. (a) The learned neural-network strength surface $\FNN=0$ (solid line) and the experimental strength data (black dots), all in the tension--tension quadrant; openness is enforced over $95^\circ$--$355^\circ$. (b) The phase-field strength surface $\mathcal{F}^\eps$ (\cref{eq:PFSS_general}) obtained upon incorporation of the learned surface $\FNN=0$ (black) with the two-term Ogden strain energy (\cref{eq:Ogden}), for varying regularization lengths. The calibration strength states are marked using red crosses. Material parameters are listed in \ref{app:materials}.}
  \label{fig:exp_sbr}
\end{figure}

The results for styrene--butadiene rubber are shown in \Cref{fig:exp_sbr}: the strength surface learned by the NN strength function from the experimental strength data is shown in \Cref{fig:exp_sbr}(a), and its PFSS in \Cref{fig:exp_sbr}(b). As for natural rubber, the strength data~\cite{hamdi2006rubber} cover only the tension--tension quadrant and are noisy and non-convex (\Cref{fig:exp_sbr}(a)), so they cannot be captured well by the commonly used analytic strength surfaces. The NN strength function is trained with the same architecture and training procedure, and openness is enforced over $95^\circ$--$355^\circ$ (16 equiangularly spaced fake points with $\mf=10$). Training is stopped at $8\times 10^3$ epochs. The learned surface closely fits the noisy, non-convex experimental data while remaining open in the enforced directions over the plotted range (\Cref{fig:exp_sbr}(a)).\\

The PFSS is obtained using~\cref{eq:PFSS_general} with the hyperelastic two-term Ogden strain energy in~\cref{eq:Ogden} (parameters in \ref{app:materials}). As for natural rubber, the calibration strength states are placed on the learned surface $\FNN=0$, along the $17$ of $64$ equiangular directions ($5.625^\circ$ spacing) lying outside the open sector, plus four states along $91^\circ$, $92^\circ$, $358^\circ$ and $359^\circ$ near the edges of the open sector, $21$ in total; the QP in~\cref{eq:qp} uses 16 far-field probes equally spaced from $95^\circ$ to $355^\circ$ (both included), here with $\mf=8$. The PFSS closely matches the learned surface up to $\eps=5\,\lch$ and remains open in the enforced directions over the plotted range (\Cref{fig:exp_sbr}(b)). Thus, the NN strength function also learns the strength data of a second elastomer, and the learned surface is accurately incorporated into the phase-field theory with a different hyperelastic strain energy, here the two-term Ogden form.\\

In summary, the same NN strength function, with the same architecture and training procedure used for the analytic surfaces in \Cref{sec:syn_results}, captures experimental strength data for different materials: data that are noisy, non-convex and confined to part of the stress space (requiring openness enforcement in compressive loadings). In each case, the learned surface is incorporated accurately into the phase-field theory across regularization lengths, with both linear-elastic and hyperelastic strain energies, owing to the high parameter dimensionality of the NN strength function. Analytic strength functions, even if they could capture the experimental data, would not yield an accurate PFSS at such regularization lengths because of their low parameter dimensionality (\Cref{sec:syn_hb,sec:syn_mc}).

\subsubsection{Learned strength surfaces in three-dimensional stress space}\label{sec:exp_3d}

\begin{figure}[!htbp]
  \centering
  \includegraphics[width=\textwidth]{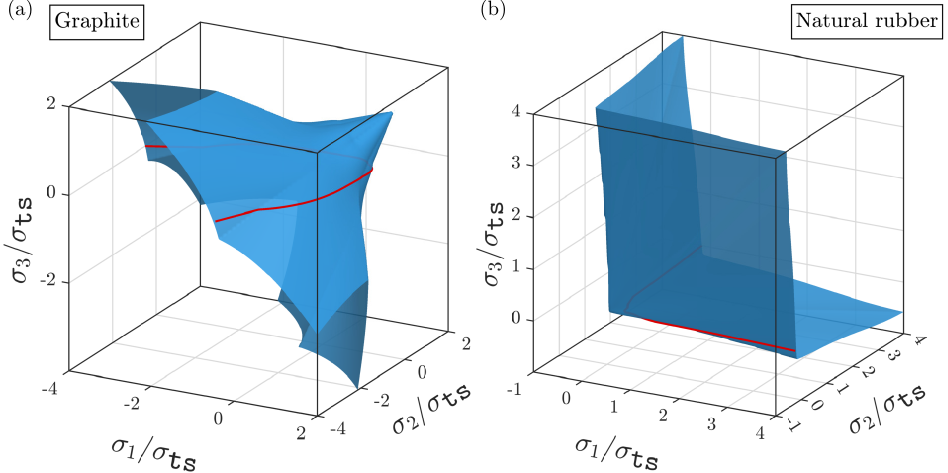}
  \caption{Learned neural-network strength surfaces $\FNN=0$ in the three-dimensional principal-stress space (normalized by $\sts$) for (a) graphite and (b) natural rubber, with the hydrostatic correction neurons of~\cref{eq:hydrofix} ($\etacrt=1.3$, $\shts=31$\,MPa for graphite; $\etacrt=1.2$, $\shts=2.9$\,MPa for natural rubber; see \Cref{tab:hydrofix}) and the compression branch switched off ($\etacrc=\phc=\infty$). The red curves are the plane-stress ($\sigma_3=0$) traces, identical to the learned surfaces in \Cref{fig:exp_graphite}(a) and \Cref{fig:exp_nr}(a).}
  \label{fig:surfaces3d}
\end{figure}

Although the training data and the learning of the NN strength surfaces have so far been restricted to the plane-stress principal-stress space, the learned NN strength surface is defined over the full stress space. Star-convexity by construction ensures physically admissible behavior of the learned surface even along loading directions outside the training space. To demonstrate this, the learned strength surfaces for graphite and natural rubber are plotted in the full three-dimensional principal-stress space in \Cref{fig:surfaces3d}(a) and (b), respectively. For these plots, the hydrostatic correction neurons in~\cref{eq:hydrofix} are enabled to prescribe the hydrostatic tensile strength $\shts$, with a cut-off $\etacrt>2/\sqrt{3}$ so that all plane-stress states are unaffected (\Cref{sec:nn_F}); the compression branch is switched off ($\etacrc=\phc=\infty$), so that no hydrostatic compressive strength is prescribed. The plane-stress ($\sigma_3=0$) traces, shown as red lines, are therefore exactly the learned strength surfaces of \Cref{fig:exp_graphite}(a) and \Cref{fig:exp_nr}(a). The hydrostatic tensile strengths are modeling inputs, since the plane-stress strength data do not constrain them. For graphite we take $\shts=31$\,MPa ($\pht=1.148$), the hydrostatic strength predicted by the Drucker--Prager surface fitted to the same graphite data~\cite{sato1987fracture}, and choose the cut-off $\etacrt=1.3$, which exceeds $2/\sqrt{3}\approx1.155$ so that the plane-stress surface is unaffected. For natural rubber we take $\shts=2.9$\,MPa ($\pht=0.0166$), the hydrostatic strength determined in~\cite{KLP21} for the natural rubber (elastomer~D, $\mu=0.588$\,MPa) of the poker-chip experiments of Gent and Lindley~\cite{GentLindley1959}, whose shear modulus is close to that of the natural rubber considered here ($\mu=0.596$\,MPa)---consistent with the weakness of elastomers in hydrostatic tension noted in Remark~\ref{rem:notconvex}~\cite{KKLP24}; here we choose $\etacrt=1.2$. The graphite surface closes to a Drucker--Prager-like apex in hydrostatic tension and remains open in hydrostatic compression. The natural rubber surface is open over most loading directions over the plotted range, and its apex in hydrostatic tension lies close to the origin, consistent with the low hydrostatic tensile strength of elastomers. In both cases, the learned surface is well behaved over the full three-dimensional principal-stress space and shows no spurious behavior beyond the training space; such robustness is essential for NN strength functions, particularly when strength data are limited.\\

{Thus, a single NN strength function learns both the widely used analytic strength surfaces and noisy, partial and non-convex experimental strength data across different materials, and its PFSS accurately reproduces the learned surface across regularization lengths, for both linear-elastic and hyperelastic strain energies, whereas the PFSS of even the exact analytic strength functions can be poor. Further, the learned strength surface remains well behaved beyond the space of the training data. In the next section, we implement the phase-field theory with the NN strength function in a finite-element framework and demonstrate its ability to model brittle fracture across the strength- and toughness-dominated regimes.}

\FloatBarrier
\section{Results: finite-element analysis with the learned strength surface}\label{sec:results_fea}

In the previous section, we established that the PFSS of the learned strength surfaces closely matches the learned surfaces across regularization lengths. These PFSS, computed from~\cref{eq:PFSS_general}, however, describe only the response to spatially uniform stress states, and they assume that the homogeneous phase field loses stability at the critical stress (\Cref{sec:pfss}). This loss of stability must be verified through numerical solution of the coupled displacement and phase-field equations. In addition, the strength-incorporated phase-field theory must preserve the Griffith physics of large cracks, which the driving force $\ce$ perturbs; that the perturbed theory follows Griffith's criticality condition with a toughness corrected by a single calibrated $\de$ is not guaranteed a priori and must be verified in boundary-value problems other than the one used to calibrate $\de$ (\Cref{sec:ce}). Finally, the theory must capture fracture nucleation governed by an interplay of strength and toughness between these two limits. The ability of the phase-field theory with the NN strength function to produce these required behaviors in boundary-value problems is therefore demonstrated in this section through finite-element analysis (FEA). Following~\cite{Chockalingam2026EFM}, we consider three canonical problems with known analytic solutions that together span the fracture regimes: (i)~strength-controlled fracture nucleation in the absence of cracks under uniform multiaxial stress (\Cref{sec:fea_uniform}), which verifies that the homogeneous phase field loses stability at the stresses given by the PFSS; (ii)~toughness-controlled nucleation and propagation from a large pre-existing crack in the double cantilever beam test (DCBT) (\Cref{sec:fea_dcbt}), which verifies that Griffith's criticality condition is recovered; and (iii)~fracture nucleation governed by an interplay of strength and toughness in the single-edge notched tension test (\Cref{sec:fea_sent}), which verifies the transition between the strength and Griffith limits as the pre-crack length is varied.\\

 We present representative results for the NN strength function learned from the experimental graphite data in \Cref{sec:exp_graphite}, which has openness enforced in part of the compression--compression quadrant. Since graphite is a linear-elastic brittle material, the strain energy throughout this section is~\cref{eq:W-mu} and the governing equations in~\cref{eq:BVP-u,eq:BVP-v} reduce to their infinitesimal-strain form. For simplicity, plane-stress conditions are assumed. The meshes are refined in the fracturing regions, with a characteristic element size $h=\eps/5$, and are coarser away from these regions. The graphite material parameters (listed in \Cref{tab:material_params}) yield a characteristic length $\lch\approx0.46$\,mm from~\cref{eq:lch}.

\runin{Implementation of the NN strength function} The simulations are performed with the open-source finite-element platform FEniCS~\cite{Logg2012FEniCS,Alnaes2015FEniCS}, using a staggered solution scheme for the displacement and phase-field equations~\labelcref{eq:BVP-u,eq:BVP-v}. The NN strength function is implemented directly in the Unified Form Language (UFL)~\cite{Alnaes2014UFL}, in which the weak forms are written in FEniCS. The invariants $r$, $\Itriax$ and $\Lode$ in~\cref{eq:r_def,eq:eta_theta} are formed as UFL expressions of the stress field, the trained network parameters enter as constants, and the layers in~\cref{eq:NN_arch} are written out as UFL expressions using the $\tanh$ and softplus ($x\mapsto\ln(1+e^{x})$) functions. $\FNN$ is then evaluated with the calibrated coefficients $\myvect{\beta}+\Dbe$ at the given $\eps$. %
 The implementation is further simplified by the form of the driving force in~\cref{eq:general_ce}. As shown in~\cite{chockalingam2025construction}, substituting~\cref{eq:general_ce} into~\cref{eq:BVP-v-main} and using $g=F+1$ gives the phase-field equation in the form
\begin{equation}
\tfrac{3}{4}\varepsilon\,\de\,G_c \triangle v_{k} = 2 v_{k} W(\bfF(\bfu_{k})) - \frac{3}{8}\,\frac{\widehat{G}^\eps_c}{\eps},\qquad \widehat{G}^\eps_c = -\de\,G_c\,F\!\left(\bm{\sigma};\myvect{\beta}+\Dbe\right),\label{eq:Gc_eff}
\end{equation}
which differs from the phase-field equation of the classical theory in two respects: the toughness $G_c$ in the gradient term is scaled by $\de$, and the toughness $G_c$ in the last term is replaced by the stress-dependent effective toughness $\widehat{G}^\eps_c$. Thus, an existing finite-element implementation of the classical phase-field theory is converted into the strength-incorporated theory by these two modifications of the weak form of the phase-field equation, which only require the evaluation of $\FNN\!\left(\bm{\sigma};\myvect{\beta}+\Dbe\right)$ described above. The values of $\de$ calibrated for each $\eps$ from the SENT test (\Cref{sec:fea_sent}), listed in \Cref{tab:de_graphite}, are used throughout this section.

\subsection{Uniform multiaxial loading: recovery of the PFSS}\label{sec:fea_uniform}

\begin{figure}[h]
  \centering
  \includegraphics[width=0.8\textwidth]{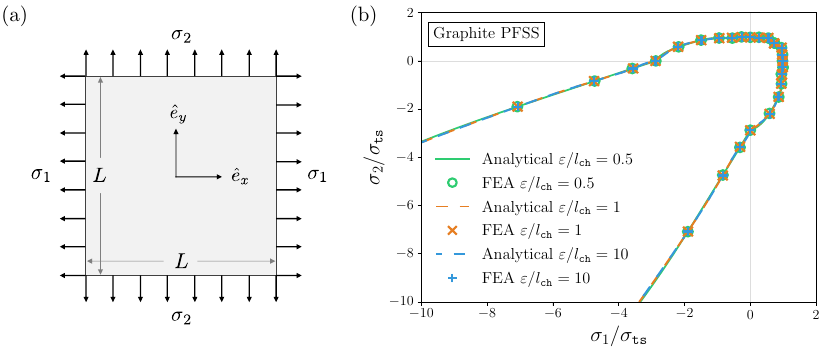}
  \caption{Uniform multiaxial loading of a square plane-stress specimen with the machine-learned strength surface for graphite (\Cref{sec:exp_graphite}). (a) Illustration of the specimen of side $L$ under uniform tractions $\sigma_1$ and $\sigma_2$. (b) The phase-field strength surface $\mathcal{F}^\eps$ (\cref{eq:PFSS_general}) of the learned graphite surface (lines) and the critical stresses from FEA (markers) along several loading directions, for varying regularization lengths.}
  \label{fig:fea_uniform}
\end{figure}

We begin with strength-controlled fracture nucleation in a pristine specimen without pre-existing cracks, subjected to a spatially uniform stress state. In this setting, the homogeneous phase field $v=1$ is expected to lose stability at the stresses defined by the PFSS in~\cref{eq:PFSS_general}~\cite{KBFLP20}; demonstrating this in FEA verifies that the PFSS computed for the learned graphite surface in \Cref{sec:exp_graphite} is the strength surface actually realized by the phase-field theory with the NN strength function. We consider a square plane-stress specimen of side $L=50\,\eps$ (\Cref{fig:fea_uniform}(a)) with face normals aligned with the principal-stress directions $\hat{e}_x$, $\hat{e}_y$~\cite{Chockalingam2026EFM}. Tractions $\tilde{\textbf{t}} = \pm\sigma_1\hat{e}_x$ and $\tilde{\textbf{t}} = \pm\sigma_2\hat{e}_y$ are applied on opposing pairs of faces. For a fixed loading direction $\hat{\bm{\sigma}}$, the magnitude is increased monotonically; the critical stress at which fracture nucleates is identified as the smallest magnitude at which the phase field $v$ first drops to zero anywhere in the specimen\footnote{This corresponds to a loss of stability of the homogeneous solution~\cite{KBFLP20,KAMAREI2026118449}. Under the present stress-controlled loading~\cite{Chockalingam2026EFM}, the phase field drops toward zero essentially uniformly. Equivalent displacement-controlled simulations, performed for selected loading directions and large specimens, were found to nucleate cracks at the same critical applied stress.}. Repeating this for many loading directions gives the FEA-predicted critical stresses in \Cref{fig:fea_uniform}(b), which are compared there with the PFSS computed from~\cref{eq:PFSS_general} and shown previously in \Cref{fig:exp_graphite}(b).\\

The FEA-predicted critical stresses agree closely with the PFSS at all regularization lengths considered. This establishes that (i) under uniform loading, the homogeneous phase field loses stability at the stresses given by the PFSS $\mathcal{F}^\eps$ of \Cref{sec:pfss}, so that the PFSS is the strength surface realized by the finite-element implementation of the phase-field theory with the NN strength function; and (ii) since the PFSS closely matches the learned surface $\FNN=0$ up to $\eps=10\,\lch$ (\Cref{sec:exp_graphite}), the finite-element implementation reproduces the learned strength surface up to $\eps=10\,\lch$. This validates a central claim of the paper end to end for the representative case of graphite.

\subsection{Double cantilever beam test: recovery of Griffith propagation}\label{sec:fea_dcbt}

\begin{figure}[h]
  \centering
  \includegraphics[width=\textwidth]{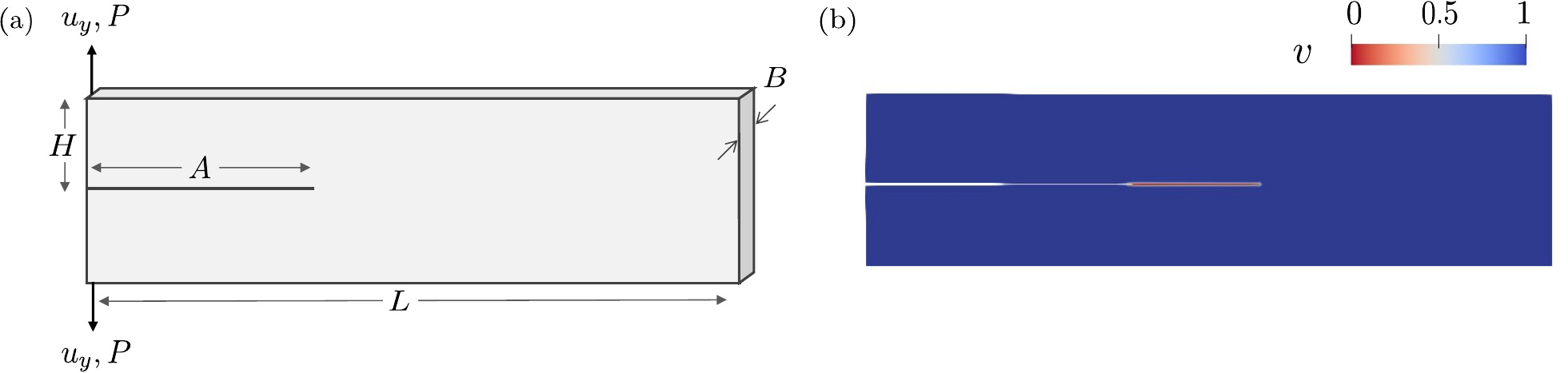}
  \caption{The double cantilever beam test (DCBT) with the machine-learned strength surface for graphite (\Cref{sec:exp_graphite}). (a) Illustration of the test: span $L$, half-thickness $H$, out-of-plane thickness $B$ and initial crack length $A$; the two arms are each displaced by $u_y$ in opposite directions, and the resulting load $P$ and crack length $a$ are recorded. (b) The crack visualized at an applied displacement $u_y=2\,u_{y,\texttt{crit}}$ through a plot of the phase field $v$ for $\eps=\lch$; the red region represents the grown crack.}
  \label{fig:fea_dcbt_geom}
\end{figure}

\begin{figure}[h]
  \centering
  \includegraphics[width=\textwidth]{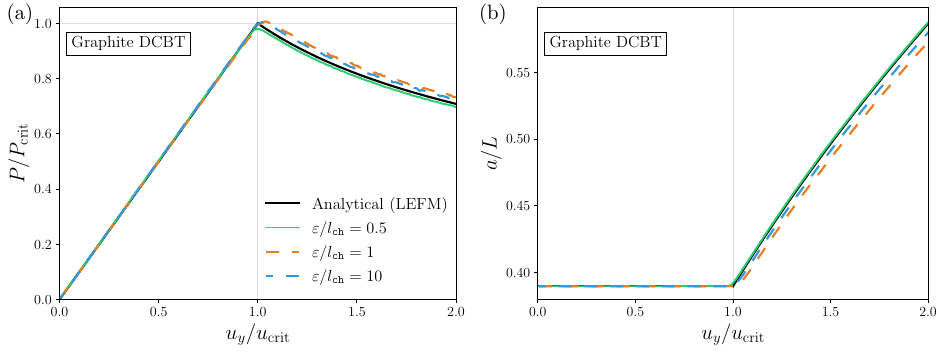}
  \caption{Results for the DCBT with the machine-learned strength surface for graphite (\Cref{sec:exp_graphite}), for varying regularization lengths. (a) Normalized load $P/P_\texttt{crit}$ and (b) normalized crack length $a/L$ as functions of the normalized applied displacement $u_y/u_{y,\texttt{crit}}$, where $P_\texttt{crit}$ is the load at $u_y=u_{y,\texttt{crit}}$. The solid black line is the linear-elastic fracture mechanics (LEFM) response in~\cref{eq:dcbt_growth,eq:dcbt_load}; the colored lines are FEA results. (a) and (b) share the same legend.}
  \label{fig:fea_dcbt_response}
\end{figure}

Next, we verify that the phase-field theory with the NN strength function preserves the Griffith physics of nucleation and propagation from large pre-existing cracks, which is controlled by the fracture toughness. This verification is essential because the driving force $\ce$ modifies the equations of the classical phase-field theory, which by themselves recover Griffith's criticality condition. Calibrating $\de$ in one test enforces agreement with the Griffith prediction only in that test; agreement in a different test, here crack propagation in the DCBT with $\de$ calibrated from the SENT test (\Cref{sec:fea_sent}), verifies that the modified theory indeed follows Griffith's criticality condition (\Cref{sec:ce}). For analytic strength surfaces, this Griffith-type behavior has been demonstrated at small regularization lengths across a variety of boundary-value problems, including mixed-mode fracture problems~\cite{KBFLP20,KDK2025Comparison,KRLP22,KKLP24,LK24,KAMAREI2026118449,Chockalingam2026EFM}. The DCBT~\cite{gilman1960direct,gillis1964double,mostovoy1967use,wiederhorn1968critical,KAMAREI2026118449} is a standard configuration for studying Mode-I crack propagation under controlled conditions and was used in~\cite{Chockalingam2026EFM} to validate the strength-incorporated phase-field theory with analytic strength surfaces. In the displacement-controlled version of this test, a pre-cracked specimen with two symmetrical arms is loaded in opening mode through applied displacement $u_y$ at the beam tips (see \Cref{fig:fea_dcbt_geom}(a)). The response load $P$ and the crack length $a$ are recorded as a function of $u_y$. The Gross--Srawley~\cite{gross1966stress} elastic analysis, combined with the Griffith criticality condition in~\cref{eq:Griffith}, yields the following closed-form expressions for the crack-growth and load--displacement responses for $L>a+3H$ and plane-stress conditions~\cite{Chockalingam2026EFM},
\begin{equation}
\textbf{Crack growth:}\quad
\begin{cases}
a = A, & u_y < u_{y,\texttt{crit}}, \\[4pt]
u_y = \dfrac{2}{\sqrt{3}}\sqrt{\dfrac{G_c H}{E}}\dfrac{a^2}{H^2}\dfrac{g_1(a/H)}{g_2(a/H)}, & u_y \ge u_{y,\texttt{crit}},
\end{cases}\label{eq:dcbt_growth}
\end{equation}
\begin{equation}
\textbf{Load--displacement:}\quad
\begin{cases}
\dfrac{P}{B} = \dfrac{E H^3}{4 A^3}\dfrac{u_y}{g_1(A/H)}, & u_y<u_{y,\texttt{crit}},\\[10pt]
\dfrac{P}{B} = \sqrt{\dfrac{G_c E H}{12}}\dfrac{H}{a}\dfrac{1}{g_2(a/H)}, & u_y\ge u_{y,\texttt{crit}},
\end{cases}\label{eq:dcbt_load}
\end{equation}
where $u_{y,\texttt{crit}}$ is the right-hand side of~\cref{eq:dcbt_growth} evaluated at $a=A$, and $g_1(y)=1+\frac{\gamma\sqrt{3}}{2y}+\frac{\gamma^2}{4 y^2}$, $g_2(y)=1+\frac{\gamma}{2\sqrt{3}\,y}$, $\gamma=2.38$.\\

We run plane-stress FEA simulations with a geometric initial crack and monotonically increasing $u_y$ up to $2\,u_{y,\texttt{crit}}$. Following~\cite{Chockalingam2026EFM}, the specimen dimensions are chosen to satisfy three requirements: (i) low beam compliance for small-deformation kinematics ($A/H=3$), (ii) dominance of large-crack Griffith physics ($A,H,L\gg\lch$), and (iii) $L>a+3H$ at the largest crack length $a$ encountered, so that~\cref{eq:dcbt_growth,eq:dcbt_load} apply. At the large regularization lengths considered here, a fourth requirement arises: (iv) the regularized crack must be sharp relative to the specimen and crack dimensions ($A,H,L\gg\eps$). This is the geometric limitation on the regularization length noted in \Cref{sec:phasefield}: the width of the regularized crack, which is of order $\eps$, must be small compared with the crack length and the beam thickness for the specimen to respond as a beam with a sharp crack, as assumed in~\cref{eq:dcbt_growth,eq:dcbt_load}. Requirements (ii) and (iv) are satisfied together by scaling the specimen dimensions with $\max(\eps,\lch)$, with half-thickness $H\approx63\max(\eps,\lch)$, $A=3H$ by (i), and $L$ satisfying (iii). The dimensions are thus fixed for $\eps\le\lch$ and grow proportionally to $\eps$ for $\eps>\lch$. ``Damaged-notch'' conditions ($v=0$ on the crack faces over a length $4\eps$ behind the initial crack tip) are imposed to avoid the geometric-crack overestimation discussed in~\cite{Tanne18,Chockalingam2026EFM}.\\

\Cref{fig:fea_dcbt_response} compares the FEA results for $\eps/\lch\in\{0.5,1,10\}$ with the linear-elastic fracture mechanics (LEFM) response in~\cref{eq:dcbt_growth,eq:dcbt_load}. \Cref{fig:fea_dcbt_response}(a) plots the normalized load $P/P_\texttt{crit}$ and \Cref{fig:fea_dcbt_response}(b) the normalized crack length $a/L$, both against the normalized applied displacement $u_y/u_{y,\texttt{crit}}$. In the LEFM response, the crack length remains at its initial value $A$ and the load increases linearly with $u_y$ up to $u_{y,\texttt{crit}}$ (elastic branch); beyond $u_{y,\texttt{crit}}$, the crack grows such that $G=G_c$ is maintained and the load decreases (fracture branch). At all regularization lengths, the FEA curves collapse onto the LEFM curves in both branches, in (a) and in (b). A representative snapshot of the phase-field solution in the deformed configuration is shown in \Cref{fig:fea_dcbt_geom}(b) at $u_y=2\,u_{y,\texttt{crit}}$ for $\eps=\lch$, where the red region is the grown crack.\\

The agreement between the FEA and LEFM responses verifies that, in the presence of large cracks, the phase-field theory with the NN strength function correctly predicts toughness-controlled fracture according to Griffith's criticality condition with the toughness $G_c$, both for the nucleation of fracture from the pre-existing crack and for its subsequent propagation, at all regularization lengths considered. Importantly, since $\de$ was calibrated from the critical stress for nucleation from a pre-crack in the SENT test (\Cref{sec:fea_sent}), a different geometry and loading, the agreement also shows that $\de$ is independent of the boundary-value problem, extending the observation of~\cite{KBFLP20,Chockalingam2026EFM} from analytic to NN strength surfaces.

\subsection{Single-edge notched tension test: the strength--toughness transition}\label{sec:fea_sent}

\begin{figure}[h]
  \centering
  \includegraphics[width=0.8\textwidth]{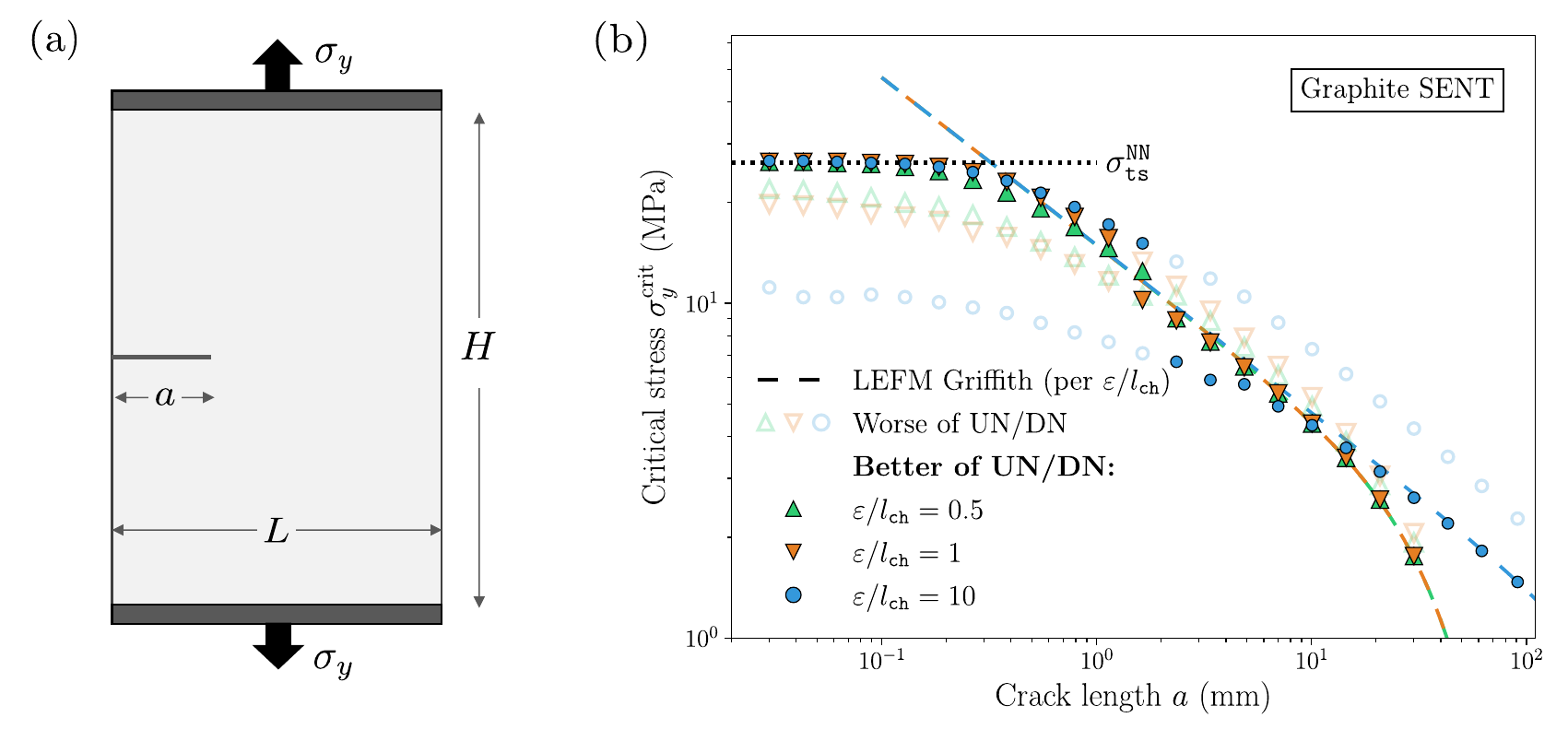}
  \caption{The single-edge notched tension (SENT) test with the machine-learned strength surface for graphite (\Cref{sec:exp_graphite}). (a) Illustration of the test. (b) Critical applied stress $\sigma_y^\texttt{crit}$ as a function of the pre-crack length $a$, for varying regularization lengths. The black dotted line is the uniaxial tensile strength $\stsNN$ of the learned surface (strength limit), and the dashed lines are the LEFM Griffith limit in~\cref{eq:sent_envelopes}, one per $\eps/\lch$ since the specimen dimensions depend on $\eps$ (\Cref{sec:fea_sent}). Markers are FEA results; for each $(\eps,a)$, the better of the undamaged-notch (UN) and damaged-notch (DN) results (closer to the analytic limits) is shown in full opacity and the other is shown faded.}
  \label{fig:fea_sent}
\end{figure}

Between the strength limit in the absence of cracks (\Cref{sec:fea_uniform}) and the Griffith limit of large cracks (\Cref{sec:fea_dcbt}), fracture nucleation is governed by an interplay of strength and toughness. To demonstrate that the phase-field theory with the NN strength function captures this interplay and transitions smoothly between the two limits, we progressively vary the pre-crack length from small to large values and show that the critical stress for fracture nucleation evolves continuously from the strength-based prediction to the Griffith prediction~\cite{Chockalingam2026EFM}. The single-edge notched tension test (SENT), illustrated in \Cref{fig:fea_sent}(a), is a standard Mode-I configuration that, when run for varying pre-crack length $a$, sweeps the entire strength--toughness transition: small $a$ probes the strength-limited regime (uniform stress in the no-crack limit) and large $a$ probes the Griffith-limited regime. The critical applied stress $\sigma_y^\texttt{crit}$ for fracture nucleation in plane stress is bounded above by the strength plateau and asymptotes to LEFM at large~$a$~\cite{Chockalingam2026EFM,tada1973stress}:
\begin{equation}
\sigma_y^\texttt{crit}(a) = \min\!\left(\stsNN,\;\;f_{\texttt{ST}}(a/L)\sqrt{\frac{E\,G_c}{\pi\,a}}\right),\label{eq:sent_envelopes}
\end{equation}
with $f_{\texttt{ST}}(a/L)$ the  SENT geometric correction factor of~\cite{tada1973stress} (see \ref{app:calibration}). Here, the strength limit is the uniaxial tensile strength of the learned surface, $\stsNN\approx26.3$\,MPa, i.e.\ the magnitude of the uniaxial tensile stress at which $\FNN=0$ (see \ref{app:materials}).\\

We run plane-stress FEA for 20 logarithmically spaced pre-crack lengths from $a=0.03$\,mm to $30$\,mm, i.e.\ from well below $\lch$ to well above, for each $\eps/\lch$. For the $\eps=10\,\lch$ case, three additional larger crack lengths up to $a=91.1$\,mm are also added. The specimen width $L$ and height $H$ (\Cref{fig:fea_sent}(a)) are  scaled with $\eps$ as $L=200\max(\eps,\lch)$ and $H=3L$, so that the specimen remains large compared to both $\lch$ and $\eps$ across the whole sweep. Since $L$ depends on $\eps$ only for $\eps>\lch$, the Griffith limits in~\cref{eq:sent_envelopes} for $\eps/\lch=0.5$ and $1$ are identical. The larger specimen for $\eps=10\,\lch$ has a different Griffith limit, which deviates visibly from the others only at large crack lengths, where $a/L$ is no longer small for the specimens with $\eps\le\lch$ ($L=91.1$\,mm, versus $911$\,mm for $\eps=10\,\lch$; \Cref{fig:fea_sent}(b)). Following~\cite{Chockalingam2026EFM}, we simulate the test under both undamaged-notch (UN) and damaged-notch (DN) conditions. Under UN conditions, the pre-crack is represented only as a geometric notch, with $v=1$ everywhere initially; under DN conditions, the phase field is additionally set to $v=0$ on the crack faces over a length $4\eps$ behind the notch tip (or over the whole crack, if shorter). The critical stress is identified as the applied stress at which $v\approx0$ at the notch tip under UN conditions, and slightly ahead of the pre-damaged region under DN conditions. The two conditions are complementary~\cite{Tanne18,Chockalingam2026EFM}. Since uniaxial tension is a calibration strength state, the UN conditions reproduce the strength limit closely at all $\eps$, but over-predict the critical stress in the Griffith limit, consistent with the known overestimation of critical loads at geometric notches in phase-field models~\cite{Tanne18}. This is because the phase-field theory recovers Griffith's criticality condition for phase-field cracks, which a geometric notch with $v=1$ is not. The DN conditions avoid this over-prediction by representing the pre-crack effectively as a phase-field crack and capture the Griffith limit closely, but under-predict the strength limit, since the pre-damaged region weakens the material; as this region scales with $\eps$, the under-prediction grows with $\eps$.  Note that in real application settings with no pre-existing large cracks, strength-based fracture nucleation and subsequent Griffith propagation of phase-field cracks would be captured accurately without the need for the damaged-notch conditions that are required here for geometric cracks.\\

\Cref{fig:fea_sent}(b) plots the critical applied stress $\sigma_y^\texttt{crit}$ against the pre-crack length $a$ for the three regularization lengths considered. The black dotted line is the strength limit $\stsNN$, the dashed lines are the Griffith limits in~\cref{eq:sent_envelopes} for each $\eps$, and the markers are the FEA results; at each $(\eps,a)$, the result of the notch condition closer to the analytic limits is shown at full opacity and the other at reduced opacity. The FEA results capture the strength limit $\stsNN$ at small $a$ and the Griffith limit at large $a$ for all three values of $\eps/\lch$. The strength--toughness transition between them is captured more accurately for $\eps/\lch=0.5$ and $1$ than for $\eps=10\,\lch$. The UN and DN results form upper and lower envelopes of the response, and the gap between these envelopes is largest in the transition region for $\eps=10\,\lch$. This sensitivity to the notch condition, and the reduced accuracy in the transition region that accompanies it, is inherent to representing a sharp geometric notch in the phase-field method and is not introduced by the NN strength function.  In the transition region for $\eps=10\,\lch$, the pre-crack length $a$ is comparable to or smaller than $\eps$, so that the regularized crack is not sharp relative to the crack length; this is the first, geometric, limitation on the regularization length noted in \Cref{sec:phasefield}, which is distinct from the strength-based limitation that the NN strength function removes. Thus, the agreement of the FEA results with the analytic strength and Griffith limits verifies that the phase-field theory with the NN strength function captures fracture nucleation in both limits across regularization lengths, and in the transition between them, governed by the interplay of strength and toughness, most accurately for the smaller regularization lengths.

\subsection{Computational implications of a large regularization length}\label{sec:fea_cost}

The results of \Cref{sec:fea_uniform,sec:fea_dcbt,sec:fea_sent} show that, with the NN strength function, the phase-field theory recovers the PFSS, Griffith propagation and the strength--toughness transition across regularization lengths, even for values as high as $\eps=10\,\lch$. We now quantify what this permits computationally. Consider a structure of characteristic dimension $\mathcal{L}$ simulated with regularization length $\eps$, and let $N_\eps=\mathcal{L}/\eps$ denote the number of regularization lengths spanned by the structure. As noted in \Cref{sec:phasefield}, the mesh size in fracturing regions must be a small fraction of $\eps$, here $h=\eps/5$, since the numerical fracture toughness exceeds $G_c$ by an amount that grows with $h/\eps$~\cite{Bourdin00,Bourdin08}; a mesh that is coarse relative to $\eps$ thus inflates the toughness and degrades the toughness-controlled response. If the crack path is not known a priori, so that the entire structure must be refined, the number of elements scales as $(\mathcal{L}/h)^2=(5N_\eps)^2$ in two dimensions and $(5N_\eps)^3$ in three dimensions; a 20-fold reduction of $\eps$ then increases the number of elements by a factor of about $400$ in two dimensions and $8000$ in three dimensions.\\

Adaptive mesh refinement, in which the mesh is refined only in the neighborhood of the evolving crack~\cite{Burke2010adaptive,Heister2015primal}, avoids refining the entire structure. Recent work has extended this to strength-controlled nucleation, using a refinement indicator based on violation of the material strength surface, combined with coarsening of the fully fractured region behind the crack tip into a coarse crack band~\cite{Dahal2026coarsening}, and has further confined the repeated solution effort to a small patch around the crack tip by condensing the inactive exterior onto the patch interface~\cite{Dahal2026localization}. A larger admissible regularization length is complementary to these strategies, as it reduces the number of elements required in every refined region.\\

\begin{figure}[h]
  \centering
  \includegraphics[width=\textwidth]{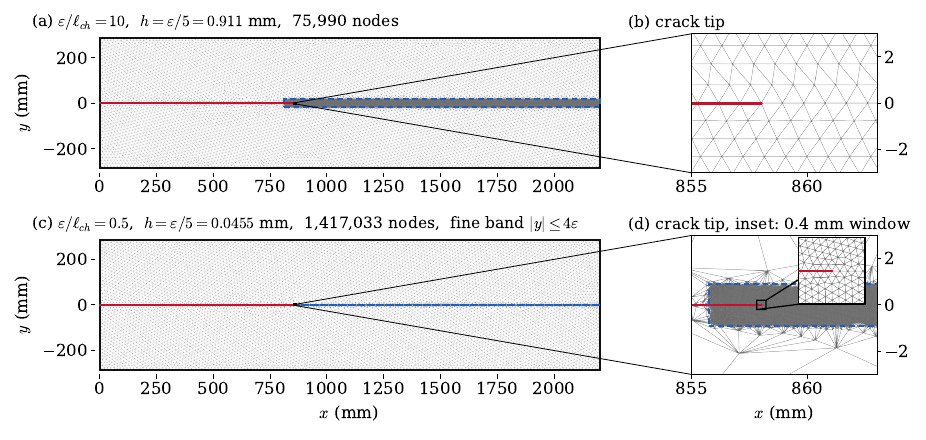}
  \caption{Finite-element meshes of the same DCBT specimen ($H=286$\,mm, $A=858$\,mm, $L=2202$\,mm; the specimen used for $\eps=10\,\lch$ in \Cref{sec:fea_dcbt}) generated with the same meshing rules for two regularization lengths. (a) Full mesh for $\eps/\lch=10$ ($h=\eps/5=0.911$\,mm, $75{,}990$ nodes) and (b) close-up of the mesh at the initial crack tip. (c) Full mesh for $\eps/\lch=0.5$ ($h=\eps/5=0.0455$\,mm, $1{,}417{,}033$ nodes) and (d) close-up at the initial crack tip, with an inset showing a $0.4$\,mm window around the tip. The red line is the initial crack and the blue dashed rectangle is the refinement zone, in which the element size is $h=\eps/5$; it extends from $10\eps$ behind the initial crack tip to the end of the specimen and over $|y|\le4\eps$ about the crack plane. The target element size outside the refinement zone is $H/20$.}
  \label{fig:fea_dcbt_mesh}
\end{figure}

Even when the crack path is known a priori, as in the DCBT, a mesh refined statically along the path must extend over a structural dimension, here the specimen length. We illustrate the typical cost on the DCBT specimen used for $\eps/\lch=10$ in \Cref{sec:fea_dcbt} for graphite, with half-thickness $H=286$\,mm, initial crack length $A=858$\,mm and span $L=2202$\,mm, meshed with the same meshing rules for $\eps/\lch=0.5$ and $\eps/\lch=10$ (\Cref{fig:fea_dcbt_mesh}). These specimen dimensions are large compared with the Irwin characteristic length of graphite, $\lch\approx0.46$\,mm; for instance, $H\approx630\,\lch$. The element size is $h=\eps/5$ in a rectangular refinement zone that extends along the crack path from $10\eps$ behind the initial crack tip to the end of the specimen and over $|y|\le4\eps$ about the crack plane, and the target element size is $H/20$ elsewhere. For $\eps/\lch=10$, the mesh has $1.51\times10^{5}$ elements ($7.6\times10^{4}$ nodes), whereas for $\eps/\lch=0.5$ it has $2.83\times10^{6}$ elements ($1.42\times10^{6}$ nodes), i.e.\ about $19$ times more. This is consistent with the number of elements in a refinement zone of width proportional to $\eps$ along a path of fixed length growing in proportion to $N_\eps$, here by the factor $20$ by which $\eps$ is reduced. (The DCBT simulations at $\eps\le\lch$ in \Cref{sec:fea_dcbt} use a correspondingly smaller specimen, since the dimensions scale with $\max(\eps,\lch)$; here the same specimen is meshed to isolate the effect of $\eps$.) This large multiplier arises even in two dimensions with a known crack path, the most favorable case; in three dimensions, the number of elements in the refinement zone around a crack surface of fixed area grows as $N_\eps^2$.\\

The displacement problem is solved here with the conjugate-gradient method preconditioned by algebraic multigrid, whose cost per linear solve grows nearly linearly with the number of unknowns $N$~\cite{Stuben2001review}; with sparse direct solvers the cost grows as $O(N^{3/2})$ in two dimensions~\cite{George1973nested,Lipton1979generalized}. The computational cost of each load step is therefore reduced at least in proportion to the number of unknowns. In three dimensions the contrast is considerably more severe: the number of elements grows faster with $N_\eps$, as discussed above, and the cost of sparse direct solvers grows as $O(N^{2})$~\cite{Lipton1979generalized}. Removing the strength-based limit on the regularization length thus makes phase-field simulation of strength-controlled fracture in large, and in particular three-dimensional, structures tractable.\\

{In summary, the finite-element simulations confirm that, with the NN strength function, the phase-field theory correctly captures fracture in all regimes across regularization lengths, even up to $\eps=10\,\lch$: strength-controlled nucleation at the stresses given by the PFSS under uniform stress states, toughness-controlled nucleation and propagation from large pre-existing cracks according to Griffith's criticality condition, and nucleation governed by the interplay of strength and toughness in the transition between these two limits. The large regularization lengths thus made accessible permit substantially coarser meshes than are required at small regularization lengths (for problems with large structural dimensions).} We summarize the main findings and outline directions for future work in the next section.

\section{Concluding remarks}\label{sec:conclusions}

In this manuscript, we introduced a neural-network (NN) strength function that learns the strength surface of isotropic brittle materials from strength data
and integrates with the strength-incorporated phase-field framework of Kumar et al.~\cite{KFLP18,KBFLP20} through the general driving-force construction of Chockalingam~(2025)~\cite{chockalingam2025construction}. The principal contributions and findings are as follows.

\begin{enumerate}
\item \emph{A star-convex-by-construction neural-network architecture for the strength function.} The strength function is constructed from a neural network that takes as input the two stress invariants that define the loading direction, the scaled triaxiality $\Itriax$ and the Lode angle $\Lode$, and outputs a positive loading-direction-dependent factor $\phihat(\Itriax,\Lode)$. Scaling $\phihat$ by the magnitude of the deviatoric stress, $r=\sqrt{J_2}/\sts$, gives the strength function $\FNN=g_{\mathtt{NN}}-1$, where $g_{\mathtt{NN}}=r\,\phihat$ is positively homogeneous of degree one in the stress, so that $\FNN$ is  star-convex with respect to the origin for any values of the network parameters; this is the primary physical requirement on a strength surface~\cite{KumarYavari2026starconvexity}. The strength function is linear in the output weights $\myvect{\beta}_{\mathtt{NN}}$, which are therefore used as the calibration parameters in the driving-force construction of Chockalingam~(2025)~\cite{chockalingam2025construction}. The positivity of $\phihat$ ensures that the strength surface is closed along every non-hydrostatic loading direction, while openness over the relevant range of stresses is enforced in chosen directions through far-field fake points during training. Since the deviatoric stress vanishes on the hydrostatic axis, finite hydrostatic strengths are captured through a clip on the scaled-triaxiality input and two fixed-weight ReLU neurons of the network that are active only beyond cut-off values of the scaled triaxiality. 

\item \emph{A single strength function for diverse strength behavior.} The same NN strength function, with the same architecture and training procedure, accurately learns a wide variety of analytic strength surfaces (through synthetic data) as well as experimental strength data of various materials. It captures the Mohr--Coulomb, Drucker--Prager, Mogi--Coulomb and 3D Hoek--Brown surfaces, which span a diverse range of physical and mathematical characteristics: open and closed surfaces, convex and non-convex surfaces, smooth surfaces and surfaces with sharp corners, dependence on and independence of the intermediate principal stress, and strength functions that are homogeneous and non-homogeneous in the stress.  For experimental data, it captures the strength of a linear-elastic brittle material (graphite) and of two hyperelastic elastomers (natural rubber and styrene--butadiene rubber), from data that are noisy, non-convex, and confined to part of the stress space, with openness enforced by choice in the remaining directions. Non-convex data of this kind are typically not represented well by the commonly used analytic strength surfaces. Further, the NN strength function eliminates the need to select an appropriate analytic strength function for each new material.

\item \emph{A generalized strength-calibration procedure.} The driving-force solution of Chockalingam~(2025) incorporates any strength function into the strength-incorporated phase-field framework of Kumar et al.~\cite{KFLP18,KBFLP20}. In doing so, it recovers $n$ independent calibration strength states exactly on the phase-field strength surface (PFSS) at any regularization length, where $n$ is the number of material parameters in which the strength function is linear. In this work, we generalize this solution to (i) a least-squares solution that admits an arbitrary number of calibration strength states (covering both overdetermined and underdetermined cases, so that the number of calibration strength states is independent of the parameter dimensionality), and (ii) a quadratic-program solution that additionally enforces openness of the PFSS in chosen directions. %

\item \emph{Accurate PFSS at finite-to-large regularization length.} The high parameter dimensionality of the NN strength function enables a PFSS match to the material strength surface up to large regularization lengths, even up to ten times the Irwin characteristic length ($\eps=10\,\lch$). This is a significant improvement over the current state of the art, in which the analytic strength function is incorporated directly and the PFSS can be poor even at regularization lengths as small as $\eps=0.5\,\lch$, even when the analytic strength function describes the material strength surface exactly. The PFSS of the surfaces learned from experimental data is also accurate, with both linear-elastic (graphite, up to $\eps=10\,\lch$) and hyperelastic (natural rubber and styrene--butadiene rubber, up to $\eps=5\,\lch$) strain energies. The poor PFSS of the analytic strength functions is due to the low dimensionality of their material parameters $\myvect{\beta}$ ($n=2$), which limits the number of independent strength states that can be calibrated \textit{exactly}, and not to the quality of their fit to the data or to the number of calibration strength states: even the exact analytic surfaces, calibrated with as many calibration strength states as the NN strength function, yield a poor PFSS. Further, the NN strength function eliminates the need for the compression correction used in earlier work~\cite{KRLP22,Chockalingam2026EFM} to improve the PFSS under compressive loadings, which introduces discontinuities in the PFSS.

\item \emph{Validation of the finite-element implementation of the phase-field theory with the NN strength function.} The finite-element implementation of the phase-field theory incorporating the NN strength function was shown to capture all the required fracture behavior across regularization lengths. Using the learned graphite surface as a model case, finite-element simulations demonstrate the correct behavior in all fracture regimes: strength-controlled nucleation under uniform multiaxial stress at the critical stresses predicted by the PFSS; toughness-controlled propagation from a large pre-existing crack in the double cantilever beam test; and nucleation governed by the interplay of strength and toughness in the single-edge notched tension (SENT) test, with a smooth transition between the strength and Griffith limits. The value of $\de$ calibrated from a single SENT simulation yields the correct critical nucleation stress across crack lengths in that test and also captures the Griffith response of the double cantilever beam test, showing that the theory perturbed by the driving force follows Griffith's criticality condition with a single correction of the toughness and extending the observation that $\de$ is independent of the boundary-value problem~\cite{KBFLP20,KAMAREI2026118449,Chockalingam2026EFM} from analytic to learned strength surfaces.

\item \emph{Practical consequence for large-structure fracture simulation.} Because the framework retains strength fidelity up to $\eps=10\,\lch$, the finite-element mesh in fracturing regions, whose size must resolve $\eps$, need no longer be tied to a small $\eps$ for accurate strength-controlled nucleation; $\eps$ then only needs to be small compared with the geometric length scales of the problem, such as the structural dimensions and the sizes of pre-existing cracks and notches. This decouples strength fidelity from mesh resolution and makes phase-field simulation of large-scale structural fracture substantially more tractable, particularly in three dimensions.
\end{enumerate}

\runin{Future directions} The present NN strength function is restricted to isotropic materials. A natural extension is to anisotropic brittle materials, such as layered rocks, cortical bone and fiber-reinforced composites, which would require additional network inputs, such as invariants of the stress formed with structural tensors that describe the material symmetry, while retaining star-convexity by construction. More broadly, machine learning could be used to capture nucleation criteria for materials with more complex behavior, such as ductile materials or poro-viscoelastic biological tissues, and to incorporate them into the phase-field theory. A complementary direction is the construction of pseudo-parameterized analytic strength functions, in which physically motivated analytic forms are augmented with additional parameters, so as to retain the physical basis of the analytic forms and their advantages, such as interpretability and reliable extrapolation beyond the available data, while providing the high parameter dimensionality that, as shown here for the NN strength function, enables an accurate PFSS at large regularization lengths. Other parameterizations of the NN strength function are also possible, for example using stress invariants that remain well defined on the hydrostatic axis as inputs, which would remove the need for the hydrostatic correction neurons; a systematic study of such alternatives, together with hyperparameter and ablation studies, was not pursued here.

\appendix
\section{Material parameters, training parameters, and calibration values}\label{app:parameters}

\subsection{Material parameters and strain-energy functions}\label{app:materials}

\Cref{tab:material_params} lists the material parameters used in this work. Indiana Limestone parameters are taken from~\cite{Chockalingam2026EFM,epp2018laboratory,schmidt1976fracture}; Titania parameters from~\cite{KBFLP20,ely1972strength,iuga2007ab}; graphite parameters from~\cite{KBFLP20,sato1987fracture,goggin1967graphite}; natural rubber (NR) and styrene--butadiene rubber (SBR) data from~\cite{hamdi2006rubber}. For graphite, the nominal tensile strength $\sts=27$\,MPa used in previous studies~\cite{KBFLP20,chockalingam2025construction} is used as the normalizing stress in $r$ in~\cref{eq:r_def}, in $\lch$ in~\cref{eq:lch}, giving $\lch\approx0.46$\,mm, and in the normalized plots in \Cref{sec:exp_graphite,sec:exp_3d,sec:fea_uniform}; the three uniaxial-tension measurements in the data range from $23.8$ to $28.5$\,MPa with mean $26.1$\,MPa, and the uniaxial tensile strength of the learned surface, $\stsNN\approx26.3$\,MPa, is close to this mean. For NR and SBR, $\sts$ is the normalizing stress used in~\cite{hamdi2006rubber}, where the failure Cauchy stresses are reported normalized by $174.56$\,MPa and $169.88$\,MPa, respectively; the normalized data are used directly here, and the uniaxial-tension measurements range from $0.83\,\sts$ to $1.09\,\sts$ for NR and from $0.92\,\sts$ to $1.07\,\sts$ for SBR.

\begin{table}[h]
\centering
\caption{Material parameters used in this work. For Indiana Limestone, $\sts=12.3$\,MPa is used for 3D H--B and $\sts=17.5$\,MPa for M--C and Mg--C~\cite{Chockalingam2026EFM}. }
\label{tab:material_params}
\setlength{\tabcolsep}{3pt}
\begin{tabular}{lcccccc}
\toprule
Material & $\sts$ & $\scs$ & $E$ & $\nu$ & $G_c$ & Strain energy \\
\midrule
Indiana Limestone & 12.3, 17.5\,MPa & 50\,MPa & 30\,GPa & 0.23 & 30\,J/m$^2$ & linear-elastic \\
Titania (TiO$_2$) & 100\,MPa & 1232\,MPa & 250\,GPa & 0.29 & --- & linear-elastic \\
Graphite & 27\,MPa & 77\,MPa & 9.73\,GPa & 0.131 & 91\,J/m$^2$ & linear-elastic \\
Natural rubber (NR) & 174.56\,MPa & --- & --- & $\to 0.5$ & --- & Yeoh \\
Styrene--butadiene rubber (SBR) & 169.88\,MPa & --- & --- & $\to 0.5$ & --- & two-term Ogden \\
\bottomrule
\end{tabular}

\end{table}

The strain-energy forms themselves are given in \Cref{sec:elasticity}: \cref{eq:W-mu} for the linear-elastic materials, the Yeoh form in~\cref{eq:Yeoh} for NR, and the two-term Ogden form in~\cref{eq:Ogden} for SBR. The corresponding parameter values, both taken from~\cite{hamdi2006rubber}, are
\begin{align}
\text{NR (Yeoh):}\quad & C_{10}=0.298\,\text{MPa},\;\; C_{20}=0.014\,\text{MPa},\;\; C_{30}=1.6\times 10^{-4}\,\text{MPa},\\
\text{SBR (Ogden):}\quad & (\mu_1,\mu_2)=(0.638,-0.025)\,\text{MPa},\;\; (\alpha_1,\alpha_2)=(3.03,-2.35).
\end{align}
The equivalent ground-state shear moduli are $\mu=2C_{10}=0.596$\,MPa for NR and, for SBR, $\mu=\tfrac{1}{2}(\mu_1\alpha_1+\mu_2\alpha_2)\approx0.996$\,MPa.

\subsection{Per-material training and calibration parameters}\label{app:calibration}

\Cref{tab:percase} summarizes per-material training-stopping epochs, numbers of training points, openness-enforcement angles, numbers of training fake points and PFSS-calibration probes and their multipliers $\mf$, and the numbers of PFSS calibration strength states. The hydrostatic correction neurons in~\cref{eq:hydrofix} do not affect any plane-stress result (\Cref{sec:nn_F}); their parameters for the three-dimensional surfaces of \Cref{fig:surfaces3d} are listed in \Cref{tab:hydrofix}, with the compression branch inactive for both materials ($\etacrc=\phc=\infty$). \Cref{tab:de_graphite} lists the SENT-calibrated values of $\de$ used for the graphite FEA results in \Cref{sec:fea_uniform,sec:fea_dcbt,sec:fea_sent} and for the graphite PFSS in \Cref{fig:exp_graphite}(b); for all other materials in \Cref{sec:results_learning}, $\de=1$ is used. For each $\eps$, $\de$ is calibrated from a single SENT simulation under damaged-notch conditions (\Cref{sec:fea_sent}) with pre-crack length $a=20\max(\eps,\lch)$, by adjusting $\de$ until the computed critical stress matches the Griffith limit in~\cref{eq:sent_envelopes}. The SENT geometric correction factor used in~\cref{eq:sent_envelopes} for the Griffith asymptote is
\begin{equation}
f_{\texttt{ST}}(a/L) = \frac{\cos(\pi a/(2L))}{\left[0.752 + 2.02\,(a/L) + 0.37\,(1-\sin(\pi a/(2L)))^3\right]\sqrt{(2L/(\pi a))\tan(\pi a/(2L))}}.\label{eq:f_st}
\end{equation}

\begin{table}[H]
\centering
\footnotesize
\caption{Per-material training and calibration parameters. $N_\texttt{data}$ is the number of training strength points $\bm{\sigma}_i$. ``train $N_\texttt{fake}$, $\mf$'' are the number of fake points $\bm{\sigma}_{\texttt{f}j}$ in~\cref{eq:loss} and the multiplier $\mf$ that sets their stress magnitude, $\lVert\bm{\sigma}_{\texttt{f}j}\rVert=\mf\max_i\lVert\bm{\sigma}_i\rVert$; ``QP $q$, $\mf$'' are the number of far-field probes $\bm{\sigma}_{\texttt{f}j}$ in~\cref{eq:Ab_qp} and the corresponding multiplier, $\lVert\bm{\sigma}_{\texttt{f}j}\rVert=\mf\max_i\lVert\bm{\sigma}_i\rVert$, used in the PFSS calibration~\cref{eq:qp}. Open angles are reported in degrees in the plane-stress principal-stress space, measured counterclockwise from the $\sigma_1$ axis ($\sigma_2=0$). $m$ is the number of calibration strength states used in the PFSS calibration system, drawn from the training set for synthetic cases and from equiangular rays on the learned surface (plus boundary anchors near the open sector; see \Cref{sec:exp_results}) for experimental cases.}
\label{tab:percase}
\setlength{\tabcolsep}{6pt}
\begin{tabular}{lcccccc}
\toprule
Material & Epochs (max/stop) & $N_\texttt{data}$ & Open angles & train $N_\texttt{fake}$, $\mf$ & QP $q$, $\mf$ & $m$\\
\midrule
3D H--B (synthetic) & $5\times 10^4$ & 32 & ---                       & ---  & ---  & 32 \\
M--C (synthetic) & $5\times 10^4$ & 32 & ---                       & ---  & ---  & 32 \\
D--P (synthetic) & $5\times 10^4$ & 25 & $191.25^\circ$--$258.75^\circ$ & 7, 10   & 7, 2.5  & 25 \\
Mg--C (synthetic) & $5\times 10^4$ & 32 & ---                       & ---  & ---  & 32 \\
Graphite        & $8\times 10^3$ & 48 &  $225^\circ$ & 1, 10   & 1, 5    & 33 \\
NR              & $9\times 10^3$ & 44 & $95^\circ$--$355^\circ$    & 16, 10   & 16, 2    & 21 \\
SBR             & $8\times 10^3$ & 52 & $95^\circ$--$355^\circ$    & 16, 10   & 16, 8    & 21 \\
\bottomrule
\end{tabular}
\end{table}

\begin{table}[H]
\centering
\caption{Parameters of the hydrostatic correction neurons in~\cref{eq:hydrofix} used in \Cref{fig:surfaces3d}.}
\label{tab:hydrofix}
\begin{tabular}{lccccl}
\toprule
Material & $\etacrt$ & $\shts$ & $\pht=\shts/\sts$ & $\etacrc,\ \phc$ & Basis for $\shts$\\
\midrule
Graphite & 1.3 & 31\,MPa & 1.148 & $\infty$ & Drucker--Prager fit to the data of~\cite{sato1987fracture}\\
NR & 1.2 & $2.9$\,MPa & 0.0166 & $\infty$ & poker-chip hydrostatic strength~\cite{KLP21}\\
\bottomrule
\end{tabular}
\end{table}

\begin{table}[H]
\centering
\caption{Calibrated $\de$ values for the graphite strength surface as a function of $\eps/\lch$, obtained from the SENT test of \Cref{sec:fea_sent}.}
\label{tab:de_graphite}
\begin{tabular}{lccc}
\toprule
& $\eps/\lch=0.5$ & $\eps/\lch=1$ & $\eps/\lch=10$\\
\midrule
$\de$ (graphite, NN strength surface) & 1.2243 & 0.9868 & 0.1842\\
\bottomrule
\end{tabular}
\end{table}

\FloatBarrier
\section*{Acknowledgements}

S.C.\ acknowledges the Lillian Gilbreth Postdoctoral Fellowship at Purdue University. A.B.T.\ gratefully acknowledges support from the U.S.\ Army Research Office under Award W911NF-24-1-0244. A.K.\ acknowledges support from the National Science Foundation, United States, through the grants CMMI-2404808 and 2541441, and the CEE-CoC Collaborative Cross-cutting seed grant from Georgia Institute of Technology.

\section*{Declaration of competing interest}
The authors declare that they have no known competing financial interests or
personal relationships that could have appeared to influence the work
reported in this paper.

\bibliographystyle{elsarticle-num}

\end{document}